\documentclass[manuscript,arxiv]{acmart}
\usepackage{hyperref}
\usepackage{makecell}
\usepackage{multirow}
\usepackage{enumitem}

\usepackage{amsmath}
\usepackage{mdframed}
\usepackage{xcolor}
\usepackage{subcaption}
\usepackage{bm}

\newmdenv[
    topline=true,
    bottomline=true,
    leftline=true,
    rightline=true,
    linecolor=blue,
    roundcorner=20pt,
    skipbelow=8pt,
    skipabove=10pt,
]{reviewbox}

\AtBeginDocument{%
  \providecommand\BibTeX{{%
    \normalfont B\kern-0.5em{\scshape i\kern-0.25em b}\kern-0.8em\TeX}}}

\newcommand{\githuburl}{https://github.com/thuhci/RingTool}
\newcommand{\githubname}{RingTool}
\newcommand{\githublink}{ \href{\githuburl}{\texttt{\githubname}} }
\newcommand{\dataurl}{https://github.com/thuhci/RingTool}
\newcommand{\dataname}{$\tau$-Ring}
\newcommand{\datalink}{ \href{\dataurl}{\texttt{\dataname}} }

\setcopyright{rightsretained}
\acmYear{2025} \acmVolume{1} \acmNumber{1} \acmArticle{1} \acmMonth{5}\acmDOI{}

\begin{document}

%%
%% The "title" command has an optional parameter,
%% allowing the author to define a "short title" to be used in page headers.
% \title{Enabling Continuous Subject-Aware Vocal Activity Sensing on Earphones}
\title{A Dataset and Toolkit for Multiparameter Cardiovascular Physiology Sensing on Rings}

%%
%% The "author" command and its associated commands are used to define
%% the authors and their affiliations.
%% Of note is the shared affiliation of the first two authors, and the
%% "authornote" and "authornotemark" commands
%% used to denote shared contribution to the research.
\author{Jiankai Tang}
\orcid{0009-0009-5388-4552}
\email{tjk24@mails.tsinghua.edu.cn}

\author{Kegang Wang}

\author{Yingke Ding}
\email{dyk21@mails.tsinghua.edu.cn}

\author{Jiatong Ji}

\author{Zeyu Wang}
\orcid{0009-0007-5048-1665}
\email{wang-zy23@mails.tsinghua.edu.cn}

\author{Xiyuxing Zhang}
\orcid{0009-0002-9337-2278}
\email{zxyx22@mails.tsinghua.edu.cn}

\author{Ping Chen}

\author{Yuanchun Shi}
\orcid{0000-0003-2273-6927}

\author{Yuntao Wang}
\authornote{Corresponding authors.}
\orcid{0000-0002-4249-8893}
\email{yuntaowang@tsinghua.edu.cn}

\affiliation{%
  \institution{Tsinghua University}
  \city{Beijing}
  \country{China}
}

%%
%% By default, the full list of authors will be used in the page
%% headers. Often, this list is too long, and will overlap
%% other information printed in the page headers. This command allows
%% the author to define a more concise list
%% of authors' names for this purpose.
% \renewcommand{\shortauthors}{Trovato and Tobin, et al.}

%%
%% The abstract is a short summary of the work to be presented in the
%% article.
\begin{abstract}

Smart rings offer a convenient way to continuously and unobtrusively monitor cardiovascular physiological signals. However, a gap remains between the ring hardware and reliable methods for estimating cardiovascular parameters, partly due to the lack of publicly available datasets and standardized analysis tools. In this work, we present $\tau$-Ring, the first open-source ring-based dataset designed for cardiovascular physiological sensing. The dataset comprises photoplethysmography signals (infrared and red channels) and 3-axis accelerometer data collected from two rings (reflective and transmissive optical paths), with 28.21 hours of raw data from 34 subjects across seven activities. $\tau$-Ring encompasses both stationary and motion scenarios, as well as stimulus-evoked abnormal physiological states, annotated with four ground-truth labels: heart rate, respiratory rate, oxygen saturation, and blood pressure.
Using our proposed RingTool toolkit, we evaluated three widely-used physics-based methods and four cutting-edge deep learning approaches. Our results show superior performance compared to commercial rings, achieving best MAE values of 5.18 BPM for heart rate, 2.98 BPM for respiratory rate, 3.22\% for oxygen saturation, and 13.33/7.56 mmHg for systolic/diastolic blood pressure estimation. The open-sourced dataset and toolkit aim to foster further research and community-driven advances in ring-based cardiovascular health sensing.
  
\end{abstract}


%%
%% Keywords. The author(s) should pick words that accurately describe
%% the work being presented. Separate the keywords with commas.

%% A "teaser" image appears between the author and affiliation
%% information and the body of the document, and typically spans the
%% page.
% \begin{teaserfigure}
%   \includegraphics[width=\textwidth]{Figures/Overview.png}
%   \caption{Seattle Mariners at Spring Training, 2010.}
%   \label{fig:teaser}
% \end{teaserfigure}

% \input{Response/overall}
% \newpage
%%
%% This command processes the author and affiliation and title
%% information and builds the first part of the formatted document.

\maketitle

\section{INTRODUCTION}
\label{sec:intro}

Smart rings have quickly grown in popularity due to their compact size, attractive design, and convenience, driving a significant increase in market demand~\cite{liang2021dualring,yu2024ring}. The global smart ring market was valued at \$340.9 million dollars in 2024 and is projected to expand substantially, reaching \$2,525.5 million by 2032 with a compound annual growth rate of 29.3\%\footnote{\url{https://www.fortunebusinessinsights.com/smart-ring-market-111418}}. Commercial products like the Oura Ring\footnote{\url{https://ouraring.com/}} and Samsung Galaxy Ring\footnote{\url{https://www.samsung.com/us/rings/galaxy-ring/}} use photoplethysmography (PPG) sensors for continuous, non-invasive cardiovascular monitoring, such as heart rate (HR) measurement and sleep stage tracking.  The anatomy of the finger uniquely enables effective sensing through its thin skin, although finger-based sensing faces distinct challenges, such as motion artifacts due to frequent hand movements. Recent studies have shown smart rings can measure other health parameters, including blood oxygen saturation (SpO2)~\cite{magno2019self}, blood pressure (BP)~\cite{osman2022blood}, and bioimpedance~\cite{usman2018ring,usman2019analyzing}, highlighting their potential for personal health management and early detection of cardiovascular diseases~\cite{wang2025computing, fiore_use_2024,kim2019towards}.

As smart ring devices gain popularity, users increasingly expect reliable monitoring of various physiological parameters across diverse scenarios, including daily activities~\cite{haveman2022continuous} and varying health conditions~\cite{breteler2025reliability}. However, comprehensive validation of ring reliability remains limited. Most studies rely on processed metrics like heart rate and sleep stage provided by commercial rings~\cite{koskimaki2018we,malakhatka2021monitoring,poongodi2022diagnosis,rajput2023assessment}, with little access to raw sensor data. This limitation restricts algorithm validation and innovation in ring-based cardiovascular sensing. Furthermore, no publicly available datasets currently provide extensive multi-parameter physiological data from wearable rings, presenting a gap in the field. The absence of standardized benchmarks further complicates direct comparisons between algorithms, restricting the robustness and general applicability of existing models. Addressing these gaps is essential for improving ring-based health monitoring technologies.

To address these limitations, we implemented two custom ring devices featuring reflective and transmissive PPG optical paths, designed to collect high-quality raw physiological signals alongside ground-truth labels. Each device integrates infrared and red-channel PPG sensors and a 3-axis accelerometer (ACC), capturing synchronized data at sampling rates up to 100 Hz. Our comprehensive data collection involved 34 subjects, with a total of 28.21 hours of recordings across various daily activities, including sitting, standing, talking, and walking. We also induced physiological variability with low oxygen saturation and elevated blood pressure through low-oxygen stimulus and deep squat exercise. The resulting $\tau$-ring dataset includes extensive raw PPG and ACC data, along with labels for blood volume pulse waveforms (BVP), respiratory waveforms (Resp), HR, respiratory rate (RR), SpO2, BP, and specific activity annotations. This dataset is the first available benchmark dedicated to ring-based cardiovascular sensing, facilitating future algorithm development and validation.

Using the $\tau$-ring dataset, we evaluate four state-of-the-art methods — ResNet~\cite{he2016resnet}, InceptionTime~\cite{ismail2020inceptiontime}, Transformer~\cite{vaswani2017transformer}, and Mamba~\cite{gu2023mamba}—across four cardiovascular monitoring tasks: HR, RR, SpO2, and BP. Our benchmarking results achieved mean absolute errors (MAE) of 5.18 BPM for HR, 2.98 breaths/min for RR, 3.22\% for SpO2, 13.48 mmHg for systolic blood pressure (SBP), and 7.47 mmHg for diastolic blood pressure (DBP). These results demonstrate performance comparable to commercial devices such as the Oura and Samsung rings. To facilitate reproducibility and future research, we provide an open-source toolkit, RingTool, which integrates both traditional physics-based signal processing methods (peak detection and spectral analysis) and supervised learning approaches. RingTool enables researchers to flexibly configure methods, select sensor channels, and apply various data filtering techniques, while also supporting the integration of additional datasets and custom algorithms.

The core contributions of this work are:
\begin{enumerate}
    \item \textbf{$\tau$-ring Dataset\footnote{Available at Kaggle: \dataurl}:} We designed and implemented two rings with reflective and transmissive optical paths, establishing the first dataset dedicated to ring-based cardiovascular sensing. The dataset includes 28.21 hours of raw PPG and ACC data from 34 subjects engaged in various activities and physiological states, with synchronized ground-truth labels.

    \item \textbf{Algorithm Performance Benchmarking:} We evaluated both physics-based and supervised algorithms for predicting HR, RR, SpO2, and BP in multiple scenarios. The results are comparable with commercial rings, bringing researchers a starting point for better algorithms.
    
    \item \textbf{RingTool Evaluation Toolkit\footnote{Available at GitHub: \githuburl}:} We open-sourced RingTool, a toolkit that integrates traditional signal processing with deep learning approaches, enhancing the reproducibility and accessibility of ring-based cardiovascular sensing research.
\end{enumerate}

\section{RELATED WORK}
\label{sec: related_works}

\begin{table}[ht]
  \centering
  \caption{Summary of Open Wearable Cardiovascular Datasets}
  \label{tab: dataset_summary}
 
  \begin{tabular}{cccccc}
  \hline
  \textbf{Dataset} & \textbf{Year} & \textbf{Subjects} & \textbf{Wearables} & \textbf{Labels} & \textbf{Tasks} \\ \hline
  \href{https://sites.google.com/site/researchbyzhang/}{TROIKA}~\cite{zhang2015troika} & 2015 & 12 & Wrist & HR & Walk, Run \\ \hline
  \href{https://physionet.org/works/ WristPPGduringexercise/}{WPPG-Exercise}~\cite{jarchi2017description} &2017 & 23 & Wrist &HR & Walk, Run, Bike \\ \hline
  \href{https://figshare.com/articles/dataset/Assessment_of_the_Fitbit_Charge_2_for_monitoring_heart_rate/5934526}{Fitbit2}~\cite{benedetto2018assessment} & 2018 & 15 & Wrist & HR & Bike \\ \hline
  \href{https://github.com/hooseok/BAMI1}{BAMI}~\cite{lee2019motion} &2018 & 24 & Wrist &HR & Stand,Walk, Run \\ \hline
  \href{https://ubicomp.eti.uni-siegen.de/home/datasets/icmi18/.}{WESAD}~\cite{schmidt2018introducing} &2018 & 17 & Wrist & HR, EDA, RR, TEMP & Sit, Stand \\ \hline
  \href{https://www.esense.io/datasets/fatigueset}{FatigueSet}~\cite{kalanadhabhatta2022fatigueset} &2018 & 12 & Ear, Wrist & HR,RR,TEMP & Bike \\ \hline  
  \href{https://ubicomp.eti.uni-siegen.de/ home/datasets/sensors19/}{DaLiA}~\cite{reiss2019deep} &2019 & 15 & Wrist &HR, EDA, RR, TEMP & Sit,Stand,Walk,Exercise  \\ \hline
  \href{https://zenodo.org/records/8142332}{EarSet}~\cite{montanari2023earset} & 2023 & 30  & Ear & HR, RR & Sit, Walk, Run\\ \hline
  \href{https://siplab.org/projects/WildPPG}{WildPPG}~\cite{meierwildppg} & 2024 & 16 & Wrist, Head, Ankle &  HR & Sit, Stand, Walk, Exercise \\ \hline
  \href{https://doi.org/10.6084/m9.figshare.27011998}{WF-PPG}~\cite{ho2025wfppg} & 2025 & 27 & Wrist,  Fingertip &   HR, BP, SpO2 & Sit \\ \hline
  $\tau$-Ring (Ours) &2025 & 34 & Finger &  HR, RR, SpO2, BP & Sit, Stand, Walk, Exercise \\ \hline  
  \end{tabular}
  
\end{table}

This section reviews related works in three key areas: ring-based cardiovascular physiology sensing in Section~\ref{sec: novel_sensing_review}, open datasets for wearable cardiovascular monitoring in Section~\ref{sec: dataset_review}, and existing cardiovascular physiology sensing toolkits in Section~\ref{sec: toolkit_review}. We highlight the limitations of current research and the need for ring-based datasets and toolkits to advance the field.

\subsection{Ring-based Cardiovascular Physiology Sensing}
\label{sec: novel_sensing_review}

Ring-based devices have become increasingly popular for continuous HR monitoring, utilizing PPG and Electrocardiograph (ECG) to provide accurate measurements. For example, ~\citet{boukhayma2021ring} developed a ring-based PPG system that demonstrated a heart rate detection accuracy of 97.87\% and inter-beat interval (IBI) errors as low as 8.1 ms. ~\citet{mahmud2018sensoring} combined PPG and an Inertial Measurement Unit (IMU) in a ring, achieving HR measurement accuracies with MAE values of 3 BPM. ~\citet{miller2022validation} validated the Oura Gen 2 Ring with other commercial wearables, showing a MAE of 0.1 $\pm$ 4.5 BPM in HR estimation. These results highlight the potential of ring-based PPG devices for HR monitoring, offering a compact and efficient alternative to traditional devices~\cite{fiore_use_2024}. However, the reliability of ring-based HR monitoring across different scenarios remains a challenge, as few studies have validated the performance during dynamic activities. This emphasizes the need for further research to explore ring-based systems in real-world scenarios. 

In addition to HR, ring-based devices have shown promise for monitoring other key cardiovascular indicators, including SpO2 and BP. ~\citet{magno2019self} introduced a self-powered ring for SpO2 monitoring using solar power and low-power processing, although its performance during dynamic activities was not evaluated. ~\citet{mastrototaro2024performance} achieved an RMSE of 2.1\% in controlled hypoxic tests with 11 volunteers. For BP monitoring, ~\citet{zheng2003ringtype} first proposed a ring-based prototype that measures PPG and ECG simultaneously to calculate BP. Subsequent studies by ~\citet{liu2019feasibility}, ~\citet{osman2022blood}, and ~\citet{kim2023firstinhuman} demonstrated that ring-based systems can provide accurate cuffless BP measurements, with MAE values for systolic BP as low as 4.38 mmHg and for diastolic BP as low as 3.63 mmHg. These advancements show that ring-based sensors are capable of accurately measuring multiple cardiovascular parameters in a compact, user-friendly form. However, current research is limited by the lack of large-scale studies on SpO2 and BP under different health conditions, and the relationship between cardiovascular signals and respiratory parameters has not been sufficiently explored.

\subsection{Open Wearable Cardiovascular Dataset}
\label{sec: dataset_review}

Open datasets have become fundamental resources for advancing cardiovascular monitoring technologies, enabling researchers to develop and validate robust algorithms for physiological signal processing~\cite{zhao2024comprehensive,tang2023mmpd,liu2024summit}. The availability of such datasets has accelerated progress in both clinical and consumer health applications by providing benchmark data for comparative evaluation~\cite{gonzalez2023benchmark}. MIMIC-III~\cite{johnson2016mimic}, one of the most comprehensive clinical datasets, contains detailed physiological waveforms and vital signs from thousands of patients in intensive care settings, offering invaluable reference data for cardiovascular monitoring research. These clinical datasets establish gold standards for algorithm development but are typically collected using specialized medical equipment in controlled environments, highlighting the need for complementary wearable datasets that capture real-world variability.

Wearable cardiovascular monitoring datasets have evolved quickly as Table~\ref{tab: dataset_summary} shows, addressing the unique challenges of portable, continuous health monitoring in daily life~\cite{miller2022validation,stone2021assessing}. Early contributions like TROIKA~\cite{zhang2015troika} focused on heart rate estimation during intense physical activities using wrist-worn PPG sensors, achieving an average absolute error of 2.34 beats per minute (BPM) during fast running at speeds of up to 15 km/h. WPPG-Exercise~\cite{jarchi2017description} expanded this approach by incorporating accelerometer and gyroscope data across diverse activities, demonstrating the importance of motion artifact compensation for improving heart rate estimation accuracy. As the field progressed, datasets like WESAD~\cite{schmidt2018introducing} and DaLiA~\cite{reiss2019deep} broadened the scope to include multiple physiological measures (HR, Electrodermal Activity (EDA), RR, temperature) for stress and activity recognition. Notably, the DaLiA dataset enabled a 31\% reduction in mean absolute error for heart rate estimation on large-scale evaluations using deep learning models, achieving improved performance compared to classical methods.

More recent datasets have explored diverse sensor placements across the body to optimize cardiovascular monitoring. EarSet~\cite{montanari2023earset} examined in-ear PPG signals to address head and facial movement artifacts, while chest-mounted sensors in datasets like WESAD~\cite{schmidt2018introducing} provided robust ECG and respiratory measurements. WildPPG~\cite{meierwildppg} captured recordings across multiple body locations, including the head and wrist, demonstrating how placement affects signal quality under real-world conditions. Hand-based measurements, primarily at the wrist and fingertips as seen in WF-PPG~\cite{ho2025wfppg}, examined dual-channel PPG signals under varying contact pressures, providing insights into PPG morphology across wrist, hand, and fingertip. 

Despite these advances in multi-location sensing and exploration of diverse body placements, a notable gap remains in comprehensive finger ring-based datasets, even though finger vessels are physiologically suitable for PPG signal acquisition. The closest open ring dataset involved only one user who self-reported health metrics (e.g., HR, Sleep Stage) measured by Oura Ring\footnote{\url{https://www.kaggle.com/datasets/eljailarisuhonen/oura-health-data-analysis-one-year-period}}. To this end, we present the $\tau$-Ring dataset, which includes 28.14 hours raw data of 34 subjects and 4 cardiovascular parameters (HR, RR, SpO2, BP) across multiple activities. This dataset aims to fill the gap in ring-based cardiovascular monitoring research and provide a valuable resource for future studies.

\subsection{Cardiovascular Physiology Sensing Toolkit}
\label{sec: toolkit_review}

Open-source toolkits play an essential role in advancing cardiovascular physiology sensing research by providing standardized methods for signal processing and analysis. Several toolkits have emerged to address specific needs in this domain, each with distinct capabilities and limitations. OmniRing~\cite{zhou2023one} represents a notable contribution in this space, providing an embedded sensor smart ring that utilizes flexible printed circuit board, 3D printing, and low-cost manufacturing technologies. PPGRaw~\cite{wolling2020quest} focuses primarily on evaluating PPG signal quality through seven decision metrics, helping researchers identify preprocessed signals and assess dataset suitability. PhysioKit~\cite{joshi2023PhysioKit} provides a standard physiological signal collection system and quality analysis function, enabling more rigorous validation of physiological sensing methods. These toolkits enable access to and systematic analysis of raw PPG data characteristics, providing valuable support for data validation.

For physiological parameter extraction, HeartPy~\cite{vangent2019heartpy,gent2019analysing}, Pyphysio~\cite{bizzego2019Pyphysio}, and NeuroKit2~\cite{makowski2021neurokit2} offer comprehensive pipelines based on traditional signal processing approaches. HeartPy specializes in heart rate analysis from PPG data, providing both time-domain and frequency-domain analysis capabilities for metrics like BPM, inter-beat interval (IBI), and various HRV indices. Its optimization for noisy signals makes it particularly useful for naturalistic settings. Pyphysio provides signal processing tools for analyzing physiological and psychological signals, especially heart rate variability, enhancing the capability for autonomic nervous system assessment. Similarly, NeuroKit2 extends these capabilities to multiple biosignals (ECG, PPG, EDA, RR), offering preprocessing, feature extraction, and visualization functions with an accessible interface for researchers.

Despite these advances, there remains a need for toolkits that support more flexible multi-channel PPG and IMU inputs, particularly for ring-form devices, while accommodating both traditional and deep learning approaches for multiple downstream tasks including HR, RR, SpO2, and BP estimation. In the remote PPG domain, frameworks like rPPG-Toolbox~\cite{liu2024rppg} and PhysBench~\cite{wang2024camera} have demonstrated how standardized tools can accelerate research progress by providing comprehensive benchmarking capabilities. rPPG-Toolbox integrates both unsupervised and supervised models with support for public datasets and systematic evaluation protocols, while PhysBench offers a benchmark framework with efficient training and HRV evaluation. In the BP measurement domain,~\citet{gonzalez2023benchmark} proposed a benchmark for cuffless BP estimation, providing a comprehensive evaluation of existing methods and datasets using PPG signals from medical devices. However, comparable toolkits designed for wearable devices, especially finger-worn sensors, are notably absent. A standard toolkit supporting multi-input, multi-task processing with both supervised and physics-based approaches would benefit the ring-based physiological sensing community.

\begin{figure}[htbp]
    \centering
    \includegraphics[width=0.9\columnwidth]{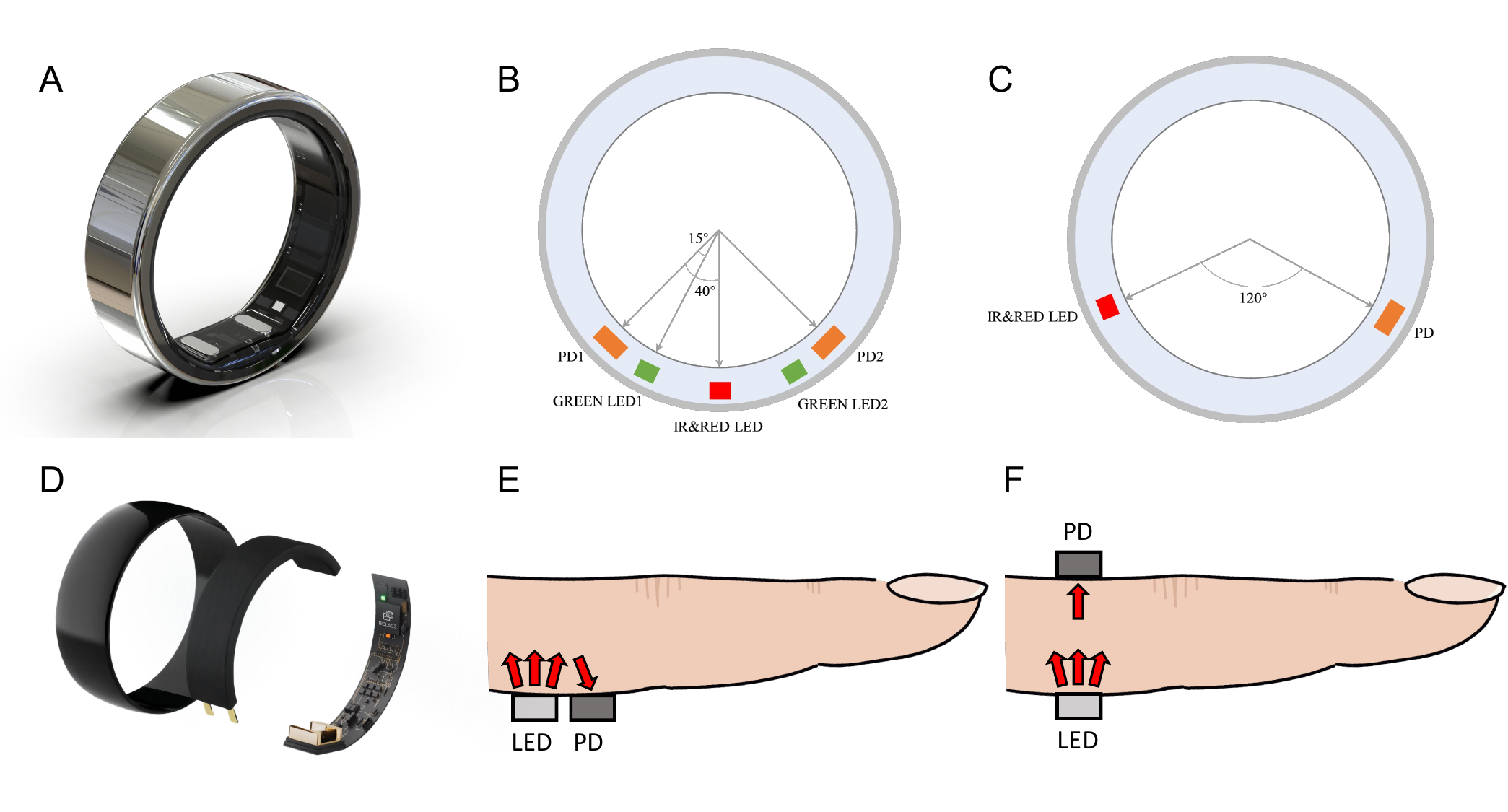}
    \caption{\textbf{Ring design and sensor layout.} This figure shows the design and sensor layout of our rings. (A) displays the physical structure of our ring. (D) shows the internal components including the battery and chip. (B) and (E) illustrate the sensor configurations for Ring 1 (reflective), showing the placement of LEDs and PDs. (C) and (F) illustrate the sensor configurations for Ring 2 (transmissive).}
    \label{fig: ring_demonstrate}
\end{figure}

\section{$\tau$-Ring DATASET}
\label{sec: dataset}

This section introduces our proposed novel dataset $\tau$-Ring. We present our designed hardware platform for synchronous data collection in Section ~\ref{sec: datasetsub1} and the collection protocol in Section ~\ref{sec: datasetsub2}. We also detail our $\tau$-Ring dataset in Section ~\ref{sec: datasetsub3} via statistical analysis. The availability and reproducibility of the $\tau$-Ring dataset are presented in Section ~\ref{sec: datasetsub4}.

\subsection{Designed Hardware Platform}
\label{sec: datasetsub1}

Inspired by OmniRing~\cite{zhou2023one}, we designed a durable, commercial-grade ring platform optimized for raw PPG and accelerometer (ACC) data collection. As shown in Figure ~\ref{fig: ring_demonstrate}A, our design features a titanium alloy outer ring and a transparent resin inner ring. The ring is lightweight at 2.5 g with compact dimensions of 8 mm in width and 2.5 mm in thickness. To ensure proper fit across participants, we manufactured rings in 8 sizes ranging from 6 to 13~\footnote{\url{https://en.wikipedia.org/wiki/Ring_size}}.

To evaluate data quality across different optical pathways, we created two ring variants with sensor layouts illustrated in Figure ~\ref{fig: ring_demonstrate}B (reflective) and Figure ~\ref{fig: ring_demonstrate}C (transmissive). Our reflective PPG ring (denoted as Ring 1) positions Light Emitting Diodes (LEDs) and photodetectors (PDs) on the same side of the finger, where light emitted by LEDs penetrates the tissue and the reflected portion is captured by adjacent PDs as shown in Figure ~\ref{fig: ring_demonstrate}E. In contrast, our transmissive PPG ring (denoted as Ring 2) places LEDs and PDs on opposite sides of the finger, allowing light to pass completely through the tissue before being detected, as illustrated in Figure ~\ref{fig: ring_demonstrate}F. This fundamental difference in optical pathway design enables us to compare signal quality and physiological measurement accuracy between reflection-based and transmission-based sensing approaches in ring form factors. 

For internal components, all rings use the nRF52840~\footnote{\url{https://www.nordicsemi.com/Products/nRF52840}} chip from Nordic as the main processing unit and a QMA6100P~\footnote{\url{https://www.qstcorp.com/en_acce_prod/QMA6100P}} three-axis accelerometer from QST. Ring 1  incorporates the GH3026~\footnote{\url{https://www.goodix.com/en/product/sensors/health_sensors/gh3026}} from Goodix supporting 3-channel PPG, while Ring 2 uses AFE4403~\footnote{\url{https://www.ti.com/product/AFE4403/part-details/AFE4403YZPR}} from Texas Instruments with 2-channel PPG capability. During data collection, we activated both infrared and red light PPG channels at a 100Hz sampling rate.
Each ring contains a battery - 18 mAh for Ring 1 (reflective) and 20 mAh for Ring 2 (transmissive). Power consumption is 2 mA for reflective mode and 5 mA for transmissive mode, allowing continuous operation throughout the data collection sessions.

\begin{figure}[htbp]
    \centering
    \begin{minipage}[b]{0.48\textwidth}
        \centering
        \includegraphics[height=0.3\textheight]{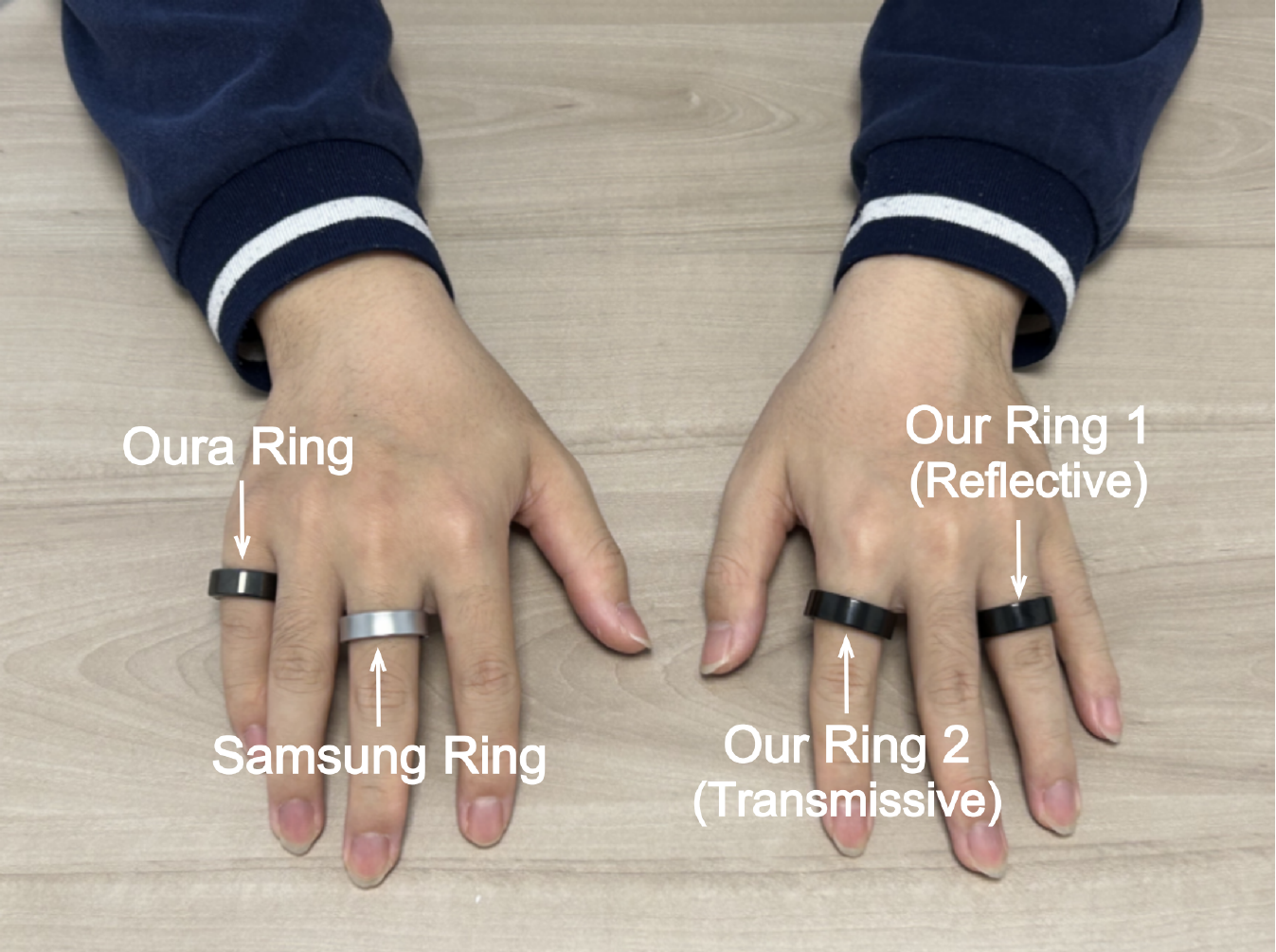}
        \caption*{(A) Users wearing our rings and commercial rings.}
        \label{fig: hand_wearing_rings}
    \end{minipage}
    \hfill
    \begin{minipage}[b]{0.48\textwidth}
        \centering
        \includegraphics[height=0.3\textheight]{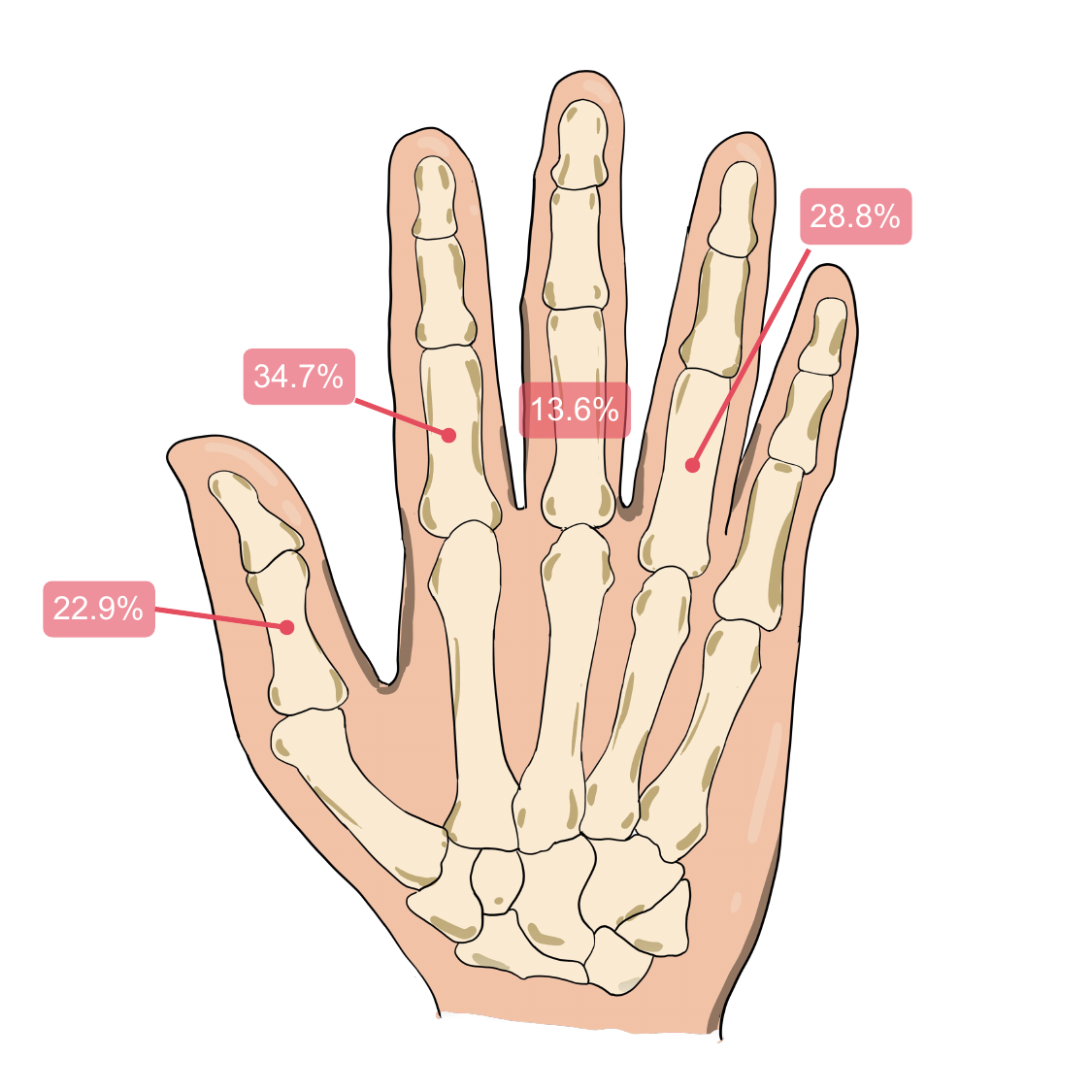}
        \caption*{(B) The position distribution of participants wearing our rings.}
        \label{fig: ring_position}
    \end{minipage}
    \caption{\textbf{Illustrations of ring usage and position distribution. } (A) shows how participants wore both our rings and commercial rings during the data collection sessions. This setup allowed for comparative analysis with established devices. (B) illustrates the distribution of positions of participants wearing our rings. }
    \label{fig: Ring position and distribution}
\end{figure}

Before each data collection session, all rings were sanitized with alcohol pads. Participants measured their finger diameter to select an appropriately sized ring, then each ring was randomly assigned to a single finger to avoid systematic bias. Our reflective and transmissive rings were worn on the one hand, while Samsung Galaxy and Oura Gen 3 rings were placed on the other hand as control devices. Figure ~\ref{fig: Ring position and distribution} shows how users worn rings and the distribution of their wearing positions with our rings. Throughout the user study, participants were required to maintain proper ring fit on their fingers to ensure consistent sensor contact and optimal data quality. During data collection, Bluetooth commands were sent via Windows-based collection software to initiate data acquisition and transmission from the rings, with Coordinated Universal Time (UTC) timestamps recorded upon each data receipt.

\subsection{Data Collection Procedure}
\label{sec: datasetsub2}

\begin{figure}[htbp]
    \centering
    \includegraphics[width=0.9\columnwidth]{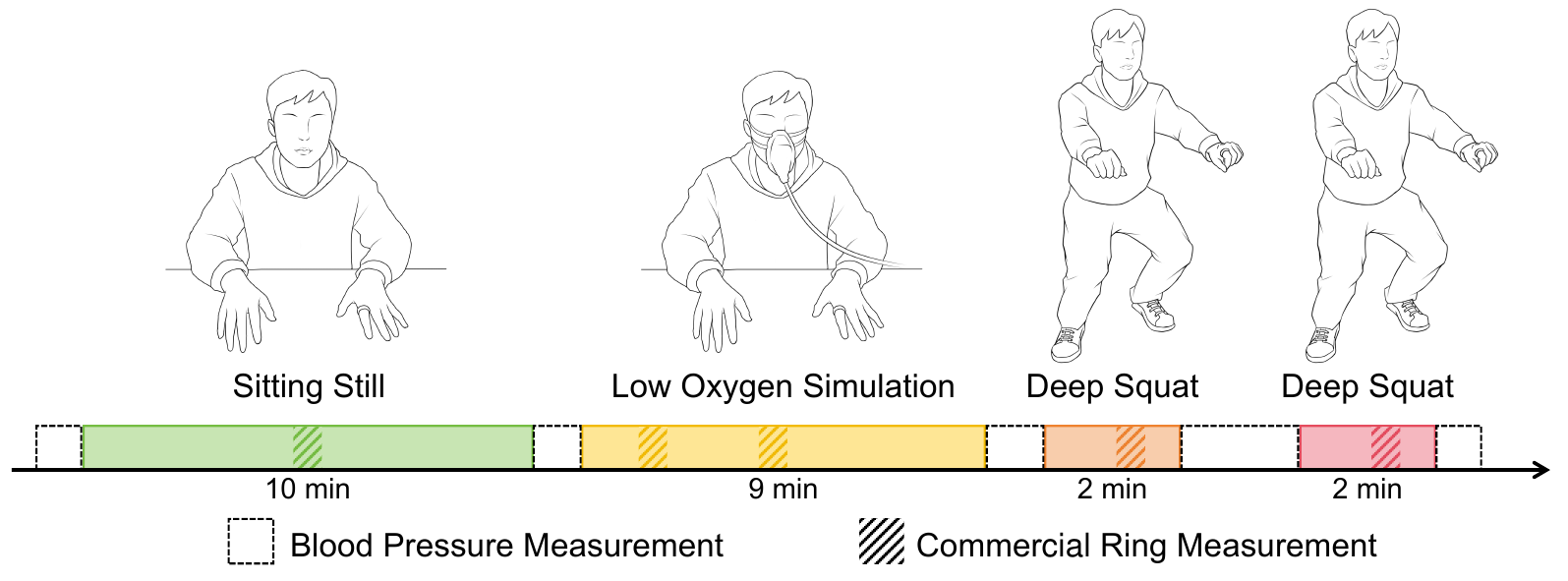}
    \caption{\textbf{Data collection procedure across stimulus-evoked physiological states.} The protocol consists of three main activities: (1) A 10-minute seated resting, (2) A 9-minute supervised low-oxygen simulation, and (3) Two 2-minute sessions of deep squat exercises. Blood pressure measurements were taken before and after each activity, while physiological data was continuously recorded by our custom rings and periodically measured by commercial rings for comparison.}
    \label{fig: health_experiment}
\end{figure}

\begin{figure}[htbp]
    \centering
    \includegraphics[width=0.9\columnwidth]{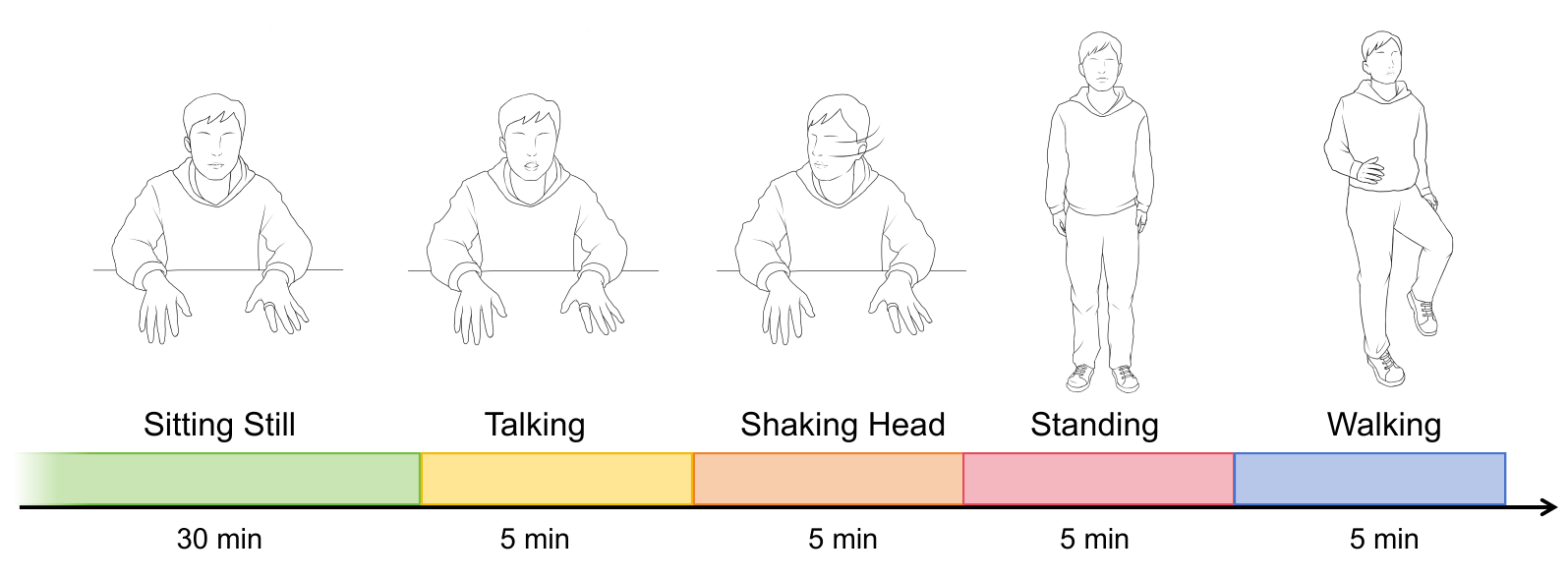}
    \caption{\textbf{Data collection procedure across daily activities.} The protocol consists of five activity segments: (1) A 30-minute seated resting, (2) 5-minute sitting and talking, (3) 5-minute head movement, (4) 5-minute standing, and (5) 5-minute walking in place. Participants wore the oximeter, Ring 1 (reflective), Ring 2 (transmissive), and respiratory band throughout all activities.}
    \label{fig: daily_experiment}
\end{figure}

In total, we recruited 34 healthy participants (20 females, 14 males) for our study. Participants independently volunteered for two separate user studies across physiological states and daily activities. The first user study (Figure ~\ref{fig: health_experiment}) included 26 participants, while the second user study (Figure~\ref{fig: daily_experiment}) involved 16 participants, resulting in a total of 42 participant sessions (8 participants voluntarily participated in both studies). All participants were recruited from the institution, including both students and staff. Their ages ranged from 18 to 30 with an average age of 21.24 (s.d. = 2.62). All participants signed a consent form prior to the collection and were informed that the collected data would be made publicly available for research purposes only. Our user study was approved by the Institutional Review Board (IRB).

\begin{figure}[htbp]
    \centering
    \begin{minipage}[b]{0.48\columnwidth}
        \centering
        \includegraphics[height=0.3\textheight]{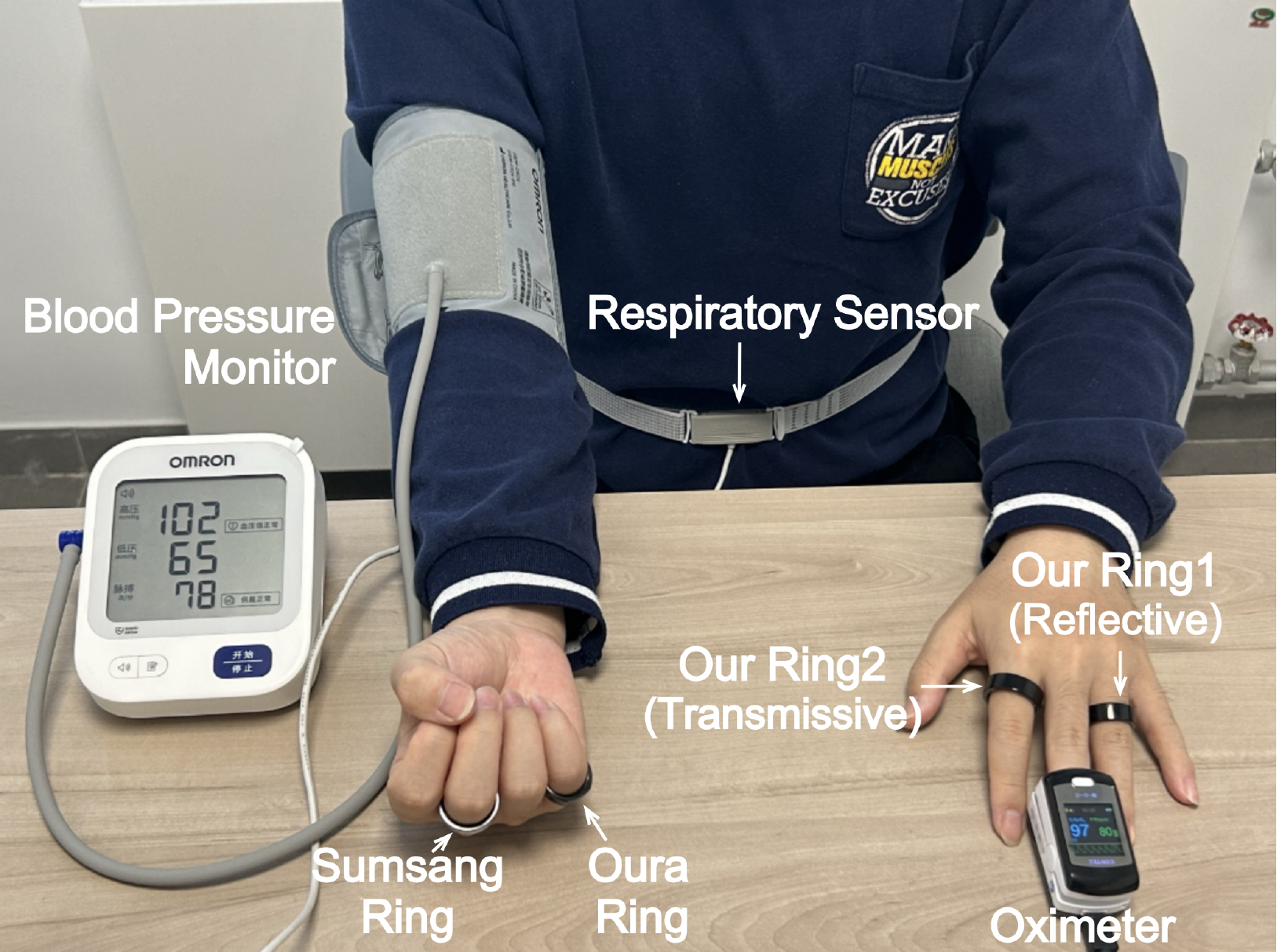}
        \caption*{(A) Participants wearing measurement devices during data collection. }
        \label{fig: equipments_wearing}
    \end{minipage}
    \hfill
    \begin{minipage}[b]{0.48\columnwidth}
        \centering
        \includegraphics[height=0.3\textheight]{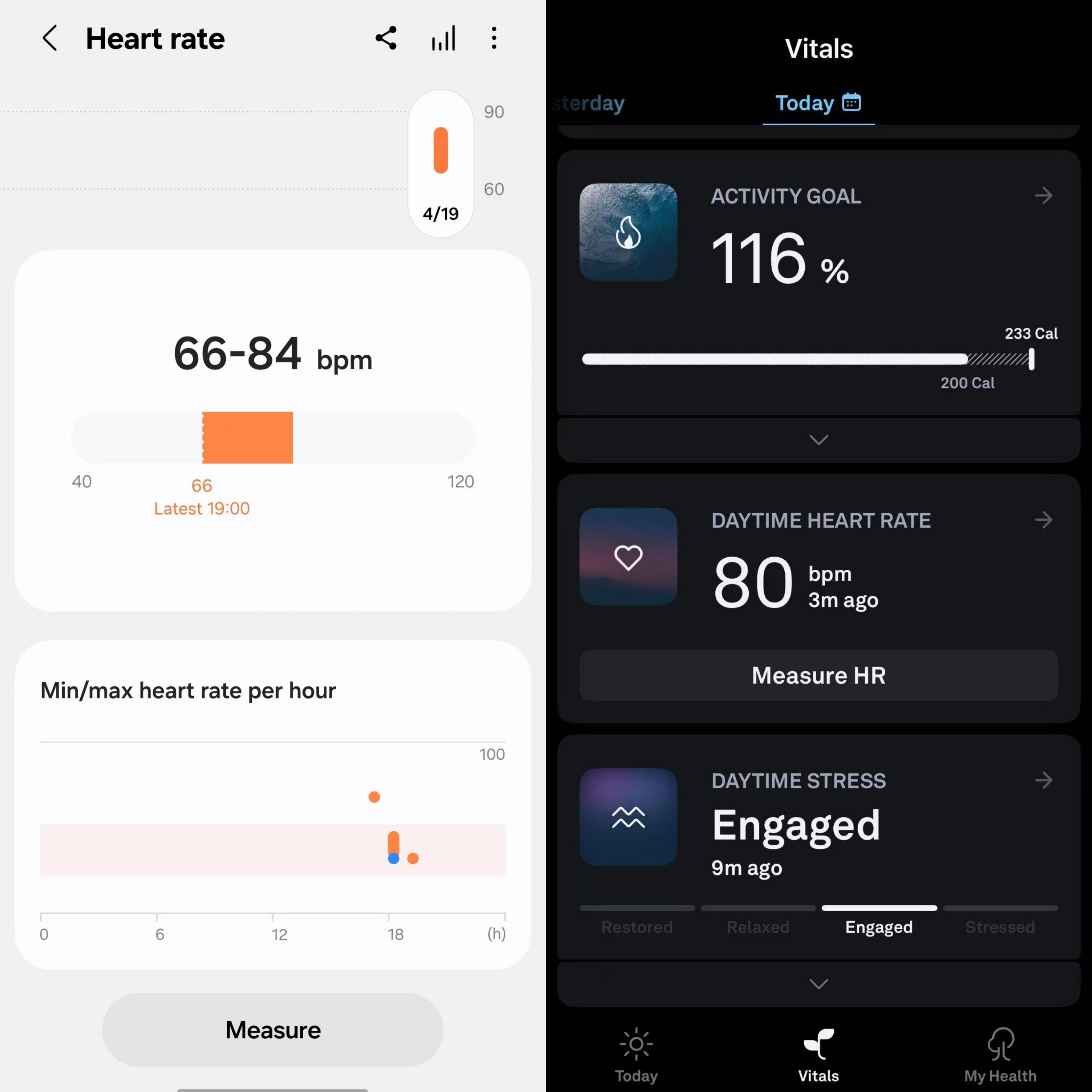}
        \caption*{(B) Measurement page of commercial rings: Samsung Galaxy Ring and Oura Gen 3. }
        \label{fig: measurement of control rings}
    \end{minipage}
    \caption{\textbf{Illustration of how we collect data during our user study. }(A) shows devices used to collect data. (B) shows the measurement of commercial rings using their apps.}
    \label{fig: combined_ring_usage}
\end{figure}

To collect ground truth physiological data, participants were equipped with a Contec CMS50E+\footnote{\url{https://www.contecmed.com/productinfo/602622.html?templateId=1133605}} oximeter, which provided blood volume pulse (BVP) at 20~Hz, HR at 1~Hz, and SpO2 at 1~Hz. Additionally, participants wore an HKH-11C+\footnote{\url{http://www.hfhuake.com/Article.asp?pid=145&id=363}} respiratory band that captured respiratory waveforms at 50~Hz. Blood pressure measurements were taken before and after activity segments using an Omron U726J\footnote{\url{https://www.omronhealthcare.com.cn/product/info/1/U726J.html}} blood pressure monitor. The wearing positions of the rings and sensors are shown in Figure ~\ref{fig: combined_ring_usage}A.

\textbf{User study across stimulus-evoked physiological states.} This user study was designed to collect physiological responses under various controlled health-related conditions, illustrated in Figure ~\ref{fig: health_experiment}. After the initial setup phase where participants were equipped with wearable rings and all monitoring devices, the protocol proceeded with three main activities: sitting, low oxygen simulation, and deep squat. Blood pressure measurements were taken before and after each activity to monitor cardiovascular changes. Participants first underwent a 10-minute resting baseline measurement in a seated position with complete stillness. Next, they completed a supervised low-oxygen simulation test consisting of three phases: 3 minutes of normal oxygen pressure, 4 minutes of reduced oxygen (10\% oxygen concentration, resulting in SpO2 dropping to 80\%-90\%~\cite{tang2025camera}), and 2 minutes of gradual recovery to normal oxygen levels. This low-oxygen protocol was approved by the IRB with participants providing explicit informed consent, as brief controlled hypoxemia induces no lasting physiological effects~\cite{bickler2017effects}. Following this, participants performed two sessions of deep squat exercises, which typically elevated heart rates to around 100 beats per minute and increased blood pressure by approximately 10~mmHg~\cite{o1962circulatory}. Throughout the user study, our custom rings continuously recorded data, while Samsung Galaxy and Oura Gen 3 rings recorded heart rate measurements in the middle of each activity with precise starting and ending measurement times for comparison purposes, as illustrated in Figure~\ref{fig: combined_ring_usage}B and Figure~\ref{fig: health_experiment}. All cardiovascular parameters were systematically documented to ensure comprehensive monitoring across the different physiological states.

\textbf{User study across daily activities.} To collect physiological data across various daily activities, we designed a systematic protocol that simulated common daily scenarios while maintaining controlled conditions, illustrated in Figure ~\ref{fig: daily_experiment}. The protocol began with a brief welcome and device setup phase, where participants were equipped with one oximeter, Ring 1 and Ring 2 on one hand, and the respiratory band around the belly. The main data collection consisted of five activity segments with precise timestamps recorded: a resting baseline where participants remained seated in complete stillness (30 minutes); sitting and talking, where participants counted aloud from 1 to 100 repeatedly (5 minutes); sitting with head movement, involving slow head rotation between 45 degrees left and right (5 minutes); standing in a comfortable posture (5 minutes); and walking in place with natural arm swinging (5 minutes). To minimize interference with normal activity patterns, measurements from blood pressure monitor and commercial ring devices were not conducted in this study.

\begin{table}[htbp]
\begin{center}
\caption{Activity and reference label details in $\tau$-Ring dataset.}
\small
\label{tab:activitycollection}
\begin{tabular}{|p{3.2cm}|p{8cm}|p{2cm}|}
\hline
\textbf{Activity} & \textbf{Description} & \textbf{Duration (hrs)} \\
\hline
All & Collective data from all activities listed below & 28.21 \\
\hline
Sitting & Participant in stationary seated position & 11.39 \\
\hline
Talking & Counting from one to one hundred repeatedly & 0.93 \\
\hline
Shaking Head & Head ratation in various directions & 0.94 \\
\hline
Standing & Participant in stationary standing position & 0.93 \\
\hline
Walking & Walking in place with natural arm swinging & 0.91 \\
\hline
Low Oxygen Simulation & Wearing mask with low-oxygen gas to simulate hypoxemia & 4.08 \\
\hline
Deep Squat & Full range of motion squat exercise & 1.61 \\
\hline
Others & User study preparation, device synchronization, blood pressure measurement and other situations not recorded with rings  & 9.26 \\
\hline
\end{tabular}

\vspace{0.5cm}

\begin{tabular}{|p{1cm}|p{5.5cm}|p{2.5cm}|p{2.5cm}|p{1cm}|}
\hline
\textbf{Label} & \textbf{Description} & \textbf{Mean$\pm$STD} & \textbf{Range (Min, Max)} & \textbf{Count} \\
\hline
BVP & Blood Volume Pulse signal samples & - & - & 1664446 \\
\hline
Resp & Respiratory waveform samples & - & - & 3953694 \\
\hline
HR & Heart Rate measurements & 80.24$\pm$12.39 BPM & (26.03, 125.93) BPM & 76586 \\
\hline
RR & Respiratory Rate calculated from Resp & 17.28$\pm$4.55 BPM & (6.04, 29.86) BPM & 4205 \\
\hline
SpO2 & Blood Oxygen Saturation measurements & 96.33$\pm$2.46 \% & (78.97, 99.00) \% & 76575 \\
\hline
SBP & Systolic Blood Pressure measurements & 106.85$\pm$16.29 mmHg & (78.00, 142.00) mmHg & 112 \\
\hline
DBP & Diastolic Blood Pressure measurements & 65.85$\pm$9.41 mmHg & (43.00, 96.00) mmHg & 112 \\
\hline
Samsung & Samsung Galaxy Ring HR measurements & 92.59$\pm$23.61 BPM & (50.00, 153.00) BPM & 128 \\
\hline
Oura & Oura Gen 3 Ring HR measurements & 89.67$\pm$22.32 BPM & (36.00, 155.00) BPM & 118 \\
\hline
\end{tabular}
\end{center}
\end{table}

\label{sec: activity_rationale}

Activities involved were strategically selected to address the gaps in existing smart ring evaluation studies. Most current research collects data only from healthy participants under resting conditions~\cite{tang2024camera,ho2025wfppg,zhang2024earsavas}, limiting the generalizability of model across daily activities and varying health states. Our dual-study approach systematically challenges ring sensors under both controlled physiological variations and common daily activities. The first study includes resting baseline measurements to establish reference values, followed by low-oxygen simulation and deep squat exercises to evaluate measurements with physiological variance. The second study complements this by capturing data during daily activities known to introduce motion artifacts and lighting fluctuations. Together, these activities create a comprehensive evaluation framework that more realistically represents the diverse conditions under which smart rings must function effectively. Table~\ref{tab:activitycollection} presents the detailed activity distribution across our dataset, highlighting our emphasis on both stationary (sitting, standing) and dynamic (walking, talking, head movement) states, as well as controlled physiological challenges (low oxygen, deep squat). Figure ~\ref{fig: visualization} visualizes the waveforms of the raw signals and reference labels, demonstrating the challenges in real-world scenarios.
This comprehensive approach addresses the notable gap in current ring-based datasets that fail to capture diverse physiological conditions across different activity states.

\subsection{Statistical Analysis of Dataset}
\label{sec: datasetsub3}

\begin{figure}[htbp]
    \centering
    \includegraphics[width=0.9\columnwidth]{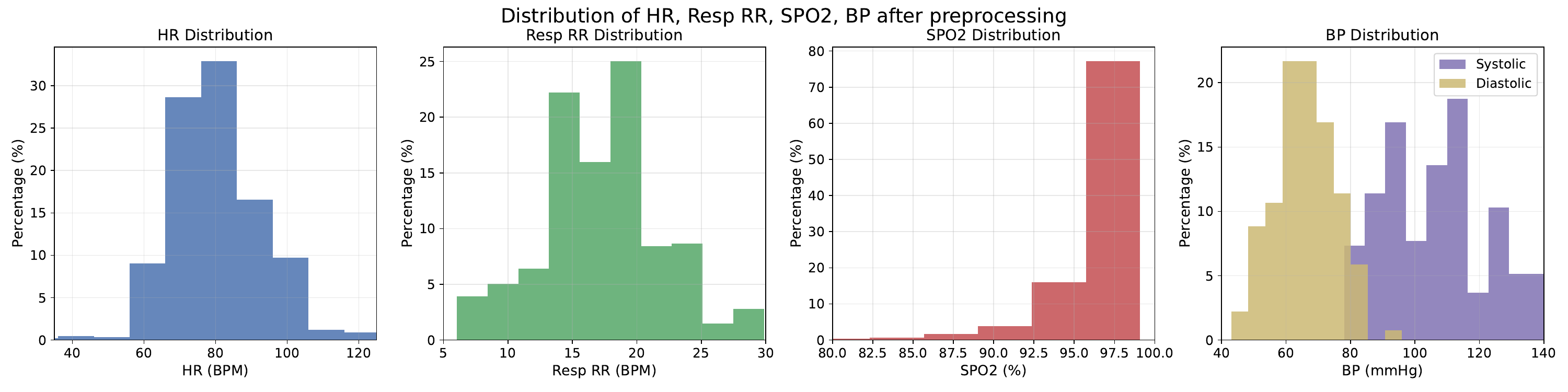}
    \caption{Ground Truth Distribution of HR, RR, SpO2, and BP.}
\label{fig: distribution}
\end{figure}

Our comprehensive data collection yielded a rich physiological dataset totaling 28.21 hours from 34 participants across 42 user studies, as detailed in Table~\ref{tab:activitycollection}. The dataset includes millions of BVP and Resp waveform samples, 76.5~K HR and SpO2 measurements, 4205 RR samples, 112 blood pressure readings and measurements from commercial rings. As illustrated in Figure~\ref{fig: distribution}, the physiological metrics demonstrate significant diversity beyond typical resting ranges, creating valuable challenges for practical ring-based sensing technology. HR measurements averaged 80.24 BPM (std: 12.39) with a wide range from 26.03 to 125.93 BPM, extending beyond typical resting values (60-100 BPM). SpO2 distribution averaged 96.33\% (std: 2.46) but notably includes values as low as 78.97\% during our controlled low-oxygen simulation, substantially deviating from normal ranges (95-100\%). Similarly, RR distribution (mean: 17.28 breaths/min, std: 4.55) ranges from 6.04 to 29.86 breaths/min, while blood pressure showed considerable variability with SBP averaging 106.85 mmHg (std: 16.26, range: 78-142 mmHg) and DBP averaging 65.85 mmHg (std: 9.39, range: 43-96 mmHg). This physiological diversity represents both a challenge and value for ring-based sensing research, as it creates a more realistic test environment addressing critical limitations in existing datasets that typically contain only normative data from subjects at rest. By capturing this spectrum of physiological states across multiple activities, our dataset enables the development of more robust algorithms capable of maintaining accuracy across real-world conditions.

\subsection{Availability of Dataset}
~\label{sec: datasetsub4}
The dataset is available via repositories on \githublink and \datalink for peer review. Currently, one fold of the five-fold dataset—comprising seven subjects used as the test set—is publicly released. Pretrained models are also provided on GitHub for reproducibility. Additional data samples and signal visualizations can be found in the GitHub repository to facilitate understanding of the data structure and quality. The complete five-fold dataset will be fully released upon acceptance.

\section{RingTool: A Toolkit to Benchmark Methods for Ring-Based Cardiovascular Metrics}
\label{sec: RingTool}

In this section, we present RingTool, a comprehensive toolkit for processing and analyzing raw PPG and ACC signals to estimate key cardiovascular parameters. RingTool offers flexible configuration options as detailed in Section~\ref{sec: configuration}, allowing researchers to customize data partitioning, training hyperparameters, and input/output specifications. The toolkit provides robust data preprocessing techniques outlined in Section~\ref{sec: data_preprocessing}, including windowing, standardization, filtering, DiffNorm, and spectral analysis to prepare raw signals for accurate parameter estimation. For traditional signal processing approaches, we implement various physics-based methods described in Section~\ref{sec: physics-based_methods}, such as peak detection, FFT analysis, and ratio-based calculations for HR, RR, and SpO2 estimation. Furthermore, RingTool incorporates supervised learning capabilities through four representative architectures (ResNet, InceptionTime, Transformer, and Mamba) with a standardized training protocol as presented in Section~\ref{sec: training_protocol}. This structured approach enables systematic evaluation and optimization of methods for ring-based cardiovascular monitoring across multiple physiological metrics and diverse usage scenarios.

\subsection{Toolkit Configuration}
\label{sec: configuration}
RingTool is built with Python 3.10, PyTorch 2.1.2 and CUDA 11.8 to process ring-based physiological data. The toolkit offers flexible configuration options across several dimensions. For dataset organization, users can divide data into training, validation, and testing sets, with support for 5-fold cross-validation. Users can adjust training parameters including learning rate, batch size, optimizer type, and training duration. Input configuration allows selection of specific sensor channels (such as different PPG wavelengths), while output configuration determines which physiological parameters to estimate (HR, RR, SpO2, BP). RingTool accommodates both physics-based signal processing methods and supervised learning approaches. The toolkit also features customizable filtering settings that can be tailored to each physiological parameter being measured. This adaptable design makes RingTool suitable for various research needs and hardware configurations.

\begin{figure}[htbp]
    \centering
    \includegraphics[width=1.0\columnwidth]{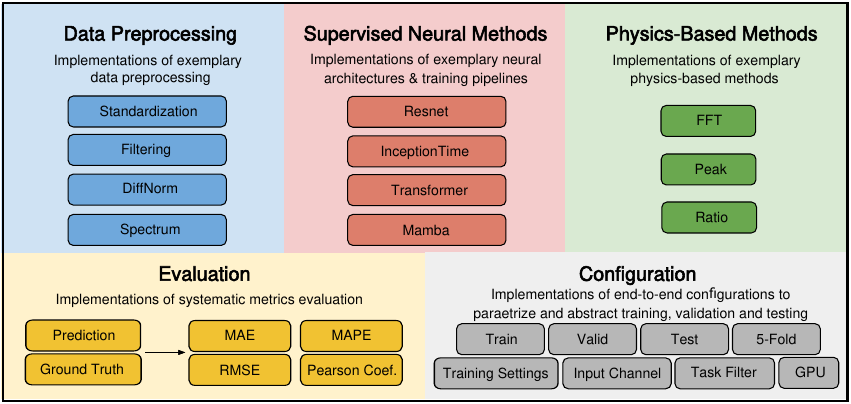}
    \caption{\textbf{Overview of Our RingTool Structure.} RingTool offers various tools for researchers to choose, including: (A) Data Preprocessing including standardization, filtering, difference-normalization and spectral analysis. (B) Supervised neural methods. (C) Physics-based methods. (D) Systematic metrics evaluations.(E) Editable configurations for dataset and training set.}
    \label{fig:structure}
\end{figure}

\subsection{Data Preprocessing}
\label{sec: data_preprocessing}
The data preprocessing pipeline for RingTool comprises five core techniques: windowing, standardization, filtering, differentiation-normalization (DiffNorm), and spectral analysis. These techniques are applied in a configurable sequence to prepare the raw signals for subsequent analysis. 

\textbf{Windowing.} Raw signals from reflective or transmissive PPG sensors are segmented using 30-second windows, filtering out segments with sampling rates below required frequency (default as 95 Hz) to ensure data quality and consistency. This duration is chosen because it encompasses multiple cardiac and respiratory cycles, enabling effective extraction of HR, RR, SpO2 and BP parameters from the captured waveforms~\cite{breteler2025reliability,haveman2022continuous}.

\textbf{Standardization.} PPG and ACC signals typically contain substantial concurrent components that vary across devices and users. We apply zero-mean unit-variance normalization to calibrate these signals, removing dimensional discrepancies and preparing the data for subsequent analysis.

\textbf{Filtering.} We employ band-pass welch filtering to isolate physiological information while removing noise. Two primary filtering configurations are implemented:
\begin{itemize}
    \item Heart Rate Extraction: Band-pass filter with frequency range 0.5-3 Hz to capture cardiac patterns
    \item Respiratory Rate Extraction: Band-pass filter with frequency range 0.1-0.5 Hz to capture respiratory patterns
\end{itemize}

\textbf{DiffNorm.} This technique involves signal differentiation followed by normalization, which enhances periodic frequency characteristics while further reducing noise components. DiffNorm has been proven particularly reliable for heart rate estimation from PPG signals by emphasizing the pulsatile components of the waveform~\cite{liu2024rppg}.

\textbf{Spectral Analysis.} Time-domain signals are transformed into the frequency domain, providing clearer visualization of dominant frequency components that correspond to physiological parameters such as heart rate and respiratory rate.

\textbf{Processing Workflow} The default workflow applies windowing, standardization, and filtering in sequence. After filtering, the signal can either undergo DiffNorm or proceed directly to spectral analysis, depending on the physiological parameter being estimated. Users can flexibly configure the sequence of these techniques according to specific requirements.

\subsection{Physics-based Methods}
\label{sec: physics-based_methods}

\textbf{Peak Detection.} We implement robust peak detection algorithms for both heart rate (30-180 beats/min) and respiratory rate (6-30 breaths/min) estimation using band-pass filtering at 0.5-3 Hz and 0.1-0.5 Hz respectively. Both parameters are computed using the same principle:

\begin{equation}
\text{Rate (per minute)} = \frac{60 \times \text{Number of peaks}}{\text{Window duration (seconds)}}
\end{equation}

\textbf{Fast Fourier Transform.} For cases where time-domain peak detection is challenged by motion artifacts, we implement frequency-domain analysis using Fast Fourier Transform (FFT). The power spectral density is computed for each signal window, and the frequency ($f_{peak}$) with maximum amplitude within the relevant physiological band is identified:

\begin{equation}
\text{Heart Rate (bpm)} = 60 \times f_{peak} \quad \text{where} \quad f_{peak} \in [0.5, 3] \text{ Hz}
\end{equation}

\begin{equation}
\text{Respiratory Rate (breaths/min)} = 60 \times f_{peak} \quad \text{where} \quad f_{peak} \in [0.1, 0.5] \text{ Hz}
\end{equation}

\textbf{Ratio-Based Methods.} For estimating blood oxygen saturation (SpO2), we utilize the same ratio calculations equation but different coefficients for different rings based on sensor configuration~\cite{severinghaus2007takuo}. The ratio is calculated as: 

\begin{equation}
R_{trans} = \frac{AC_{red}/DC_{red}}{AC_{infrared}/DC_{infrared}}
\end{equation}

SpO2 is then estimated through empirical calibration:

\begin{equation}
\text{SpO2 (\%)} = a - b \times R
\end{equation}

where $a$ and $b$ are calibration coefficients (typically $a \approx 99$ and $b \approx 6$ for reflective PPG sensors, and $a \approx 87$ and $b \approx -6$ for transmissive).

\subsection{Supervised Methods and Training Protocol}
\label{sec: training_protocol}

Inspired by WildPPG~\cite{meierwildppg}, RingTool incorporates four representative deep learning architectures that serve as common backbones for time-series physiological signal analysis. Each architecture is designed to effectively process multi-channel inputs from ring sensors while addressing the regression challenges inherent in physiological parameter estimation.

\textbf{ResNet~\cite{he2016resnet}.} Our implementation adapts the classic residual network architecture for physiological time-series analysis. The temporal ResNet processes multi-channel inputs (PPG, ACC) through parallel convolutional pathways that preserve sensor-specific information while enabling cross-channel learning. The residual connections address the vanishing gradient problem, making this architecture suitable for deeper networks needed in complex regression tasks like blood pressure estimation. Key adjustable parameters include network depth, filter counts, kernel sizes, and regularization strength, allowing optimization for different physiological signals with varying temporal characteristics. Our default ResNet implementation uses approximately 1.54M parameters.

\textbf{InceptionTime~\cite{ismail2020inceptiontime}.} This architecture implements a multi-scale approach designed for capturing patterns across different temporal resolutions simultaneously. The parallel convolutional filters with varying kernel sizes make InceptionTime especially suited for detecting both rapid cardiac events and slower respiratory cycles within the same multi-channel input. This backbone demonstrates decent performance in tasks requiring sensitivity to diverse frequency components. The architecture can be customized through filter counts, inception module depth, and kernel size distributions to accommodate specific physiological monitoring requirements. Our default InceptionTime implementation uses approximately 0.15M parameters.

\textbf{Transformer~\cite{vaswani2017transformer}.} Our transformer implementation leverages self-attention mechanisms to identify complex relationships between different timepoints and across multiple input channels. This architecture helps capture capturing long-range dependencies in physiological signals, making it useful for parameters that exhibit temporal variations across extended windows, such as respiratory rate during activity transitions. The multi-head attention design enables simultaneous focus on different physiological aspects across channels, while the encoding structure can be adjusted through attention heads, embedding dimensions, and layer count configurations. Our default Transformer implementation uses approximately 0.38M parameters.

\textbf{Mamba~\cite{gu2023mamba}.} As the newest backbone in our toolkit, Mamba represents a recent advancement in sequence modeling with its selective state space mechanism. This architecture processes multi-channel physiological inputs with linear computational scaling, making it efficient for analyzing extended signal sequences. Mamba's selective attention to signal components may provide improved handling of motion artifacts common in ring sensor data, making it applicable for continuous monitoring during physical activities. The model's state dimension, block count, and expansion factors can be configured to balance efficiency and performance across different physiological regression tasks. Our default Mamba implementation uses approximately 1.76M parameters.

\textbf{Training Protocol.} RingTool employs a standardized training framework adaptable to various physiological parameter estimation tasks. The protocol supports single-task learning for different parameters (HR, RR, SpO2, BP). Training follows a 5-fold cross-validation approach, where each fold is sequentially assigned to test folds to ensure each subject appears in the test set once.  The default configuration uses 200 epochs with a batch size of 128, Mean Squared Error (MSE) loss function, and Adam optimizer with a learning rate of 1e-3. Model selection is performed by monitoring validation set performance, with the best-performing epoch's weights being preserved. All models support optimization through common hyperparameters including learning rate, batch size, optimizer selection, and regularization techniques, with automated early stopping and model checkpointing to preserve optimal configurations.

\section{Results and Findings}
\label{sec: RingTool_Results}

\begin{table}[htbp]
    \centering
    \caption{Best results on HR, RR, SpO2, BP tasks.}
    \label{tab: all_performance}
    \begin{tabular}{r|c|cccc|c|cccc}
    \hline
    \textbf{Task} &  \multicolumn{5}{c|}{\textbf{Ring 1 (Reflective)}} & \multicolumn{5}{c}{\textbf{Ring 2 (Transmissive)}} \\
    &  \textbf{Method} & \small MAE & \small RMSE & \small MAPE & \small Pearson & \textbf{Method} & \small MAE & \small RMSE & \small MAPE & \small Pearson \\
    \hline
    \hline

    \textbf{HR} &  \small ResNet & \small $5.18$ & \small $8.96$ & \small $6.99$ & \small $0.73$ & \small ResNet & \small $8.09$ & \small $11.51$ & \small $10.49$ & \small $0.37$ \\
    \textbf{RR} & \small Peak & \small $2.98$ & \small $4.12$ & \small $18.39$ & \small $0.50$ & \small Mamba & \small $3.30$ & \small$4.28$ & \small $21.20$ & \small $0.13$  \\
    \textbf{SpO2} & \small ResNet & \small $3.22$ & \small $4.46$ & \small $3.57$ & \small $0.01$ & \small Ratio & \small $3.31$ & \small $3.95$ & \small $3.58$ & \small $0.28$ \\
    \textbf{SBP} & \small Transformer & \small{$13.33$} & \small $16.10$ & \small $12.53$ & \small $0.25 $  & \small Mamba & \small $14.56$ & \small $17.76$ & \small $13.79$ & \small $0.09$ \\
    \textbf{DBP}  & \small Transformer & \small $7.56$ & \small $9.80$ & \small $11.80$ & \small $-0.02$  & \small Transformer & \small $8.06$ & \small $10.54$ & \small $12.57$ & \small $-0.15$ \\

    \hline
    \end{tabular}
    \\
    \footnotesize{For all cardiovascular parameters: MAE = Mean Absolute Error (HR/RR: BPM, BP: mmHg, SpO2: \%), RMSE = Root Mean Square Error (HR/RR: BPM, BP: mmHg, SpO2: \%), MAPE = Mean Absolute Percentage Error (\%), Pearson = Pearson Correlation Coefficient}
\end{table}

In this section, we present comprehensive evaluation results from our benchmark methods for various cardiovascular parameters using ring-based sensing technology. We first examine HR estimation performance in Section~\ref{sec: heart_rate_estimation}, comparing different algorithms across stationary (i.e., sitting, talking, shaking head, standing and low oxygen simulation) and motion (i.e., walking and deep squat) scenarios. Next, we analyze RR estimation accuracy in Section~\ref{sec: respiratory_rate_estimation}, followed by SpO2 assessment in Section~\ref{sec: blood_oxygen_saturation_estimation}. Finally, we evaluate BP estimation capabilities in Section~\ref{sec: blood_pressure_estimation}. For each parameter, we compare both physics-based and supervised learning approaches, examining the performance differences between reflective (Ring 1) and transmissive (Ring 2) sensing modalities across various user activities and physiological states. The best results for each task are summarized in Table~\ref{tab: all_performance}. We also provide detailed performance metrics for each method in Tables~\ref{tab: hr_performance}, \ref{tab: rr_performance}, \ref{tab: spo2_performance}, and \ref{tab: sbp_performance}.

\subsection{Settings and Evaluation Metrics }
Our cardiovascular parameter estimation tasks (HR, RR, SpO2, BP) are framed as time series regression problems, where the ground-truth labels are paired with the corresponding signal sequences based on overlapping time intervals. This pairing ensures that the labels align with the input signals without abrupt changes in cardiovascular parameters. The dataset is divided into five folds, with each fold used for testing while the remaining four folds are used for training and validation. This cross-validation process is repeated five times, and the results are averaged to compute the final performance metrics. Both training and testing are conducted on the same device (Ring 1 or Ring 2) to guarantee consistency in the comparison. The training process leverages data from all scenarios in the training set, while evaluation is performed on specific scenarios from the test set. Data collected concurrently from the oximeter, our prototype rings, and commercial rings are used for comparisons with commercial rings.
After calculating and merging five-fold results, we evaluate all methods using the following standardized metrics:

\begin{itemize}
    \item Mean Absolute Error (MAE): The average absolute difference between predicted values and reference values, calculated as $\text{MAE} = \frac{1}{n}\sum_{i=1}^{n}|y_i - \hat{y}_i|$, where $y_i$ is the ground truth and $\hat{y}_i$ is the predicted value.
    
    \item Root Mean Square Error (RMSE): Measures the square root of average squared estimation errors, calculated as $\text{RMSE} = \sqrt{\frac{1}{n}\sum_{i=1}^{n}(y_i - \hat{y}_i)^2}$.
    
    \item Mean Absolute Percentage Error (MAPE): Expresses error as a percentage of the true value, calculated as $\text{MAPE} = \frac{100\%}{n}\sum_{i=1}^{n}|\frac{y_i - \hat{y}_i}{y_i}|$.
    
    \item Pearson Correlation Coefficient: Measures the linear correlation between predicted and reference values, ranging from -1 to 1, calculated as $r = \frac{\sum_{i=1}^{n}(y_i-\bar{y})(\hat{y}_i-\bar{\hat{y}})}{\sqrt{\sum_{i=1}^{n}(y_i-\bar{y})^2\sum_{i=1}^{n}(\hat{y}_i-\bar{\hat{y}})^2}}$.
\end{itemize}

\begin{table}[htbp]
\centering
\caption{Benchmark results of different methods on HR task.}
\setlength{\tabcolsep}{3pt}  
\label{tab: hr_performance}
\begin{tabular}{r|cccc|cccc}
\hline
\textbf{HR} & \multicolumn{4}{c}{\textbf{Ring 1 (Reflective)}} & \multicolumn{4}{c}{\textbf{Ring 2 (Transmissive)}} \\
\textbf{All}&  \small MAE & \small  RMSE &  
\small  MAPE & \small  Pearson &  \small MAE & \small  RMSE &  
\small  MAPE & \small  Pearson \\
\hline
\hline
\textbf{Physics-based} & & & & & & & &  \\
\small Peak & \small $18.47\pm0.44$ & \small $26.72\pm5.13$ & \small $24.72\pm0.67$ & \small $0.32\pm0.02$ & \small $28.53\pm0.51$ & \small $36.03\pm6.01$ & \small $38.24\pm0.77$ & \small $0.00\pm0.02$  \\
\small FFT & \small $9.32\pm0.40$ & \small $19.80\pm4.81$ & \small $10.90\pm0.43$ & \small $0.36\pm0.02$ & \small $15.10\pm0.46$ & \small $24.85\pm5.25$& \small $17.93\pm0.51$ & \small $0.08\pm0.02$   \\
\textbf{Supervised} & & & & & & & &  \\
\small ResNet & \small \bm{$5.18\pm0.17$} & \small  \bm{$8.96\pm2.56$} & \small \bm{$6.99\pm0.31$} & \small \bm{$0.73\pm0.02$} & \small \bm{$8.09\pm0.19$} & \small $11.51\pm2.66$ & \small \bm{$10.49\pm0.31$} & \small \bm{$0.37\pm0.02$} \\
\small InceptionTime & $6.69\pm0.18$ & \small $10.31\pm2.57$ & \small $8.75\pm0.30
$ & \small$0.62\pm0.02$ &\small $8.20\pm0.18$ & \small \bm{$11.16\pm2.42$} & \small $10.65\pm0.28$ & \small $0.36\pm0.02$ \\
\small Transformer & \small $7.94\pm0.18$ & \small $11.22\pm2.60$ & \small $10.22\pm0.33$ & \small $0.54\pm0.02$ & \small $8.70\pm0.18$ & \small $11.53\pm2.46$ & \small $11.32\pm0.29$ & \small $0.21\pm0.02$ \\
\small Mamba & \small $7.64\pm0.18$ & \small $11.05\pm2.71$ & \small $10.27\pm0.33$ & \small $0.58\pm0.02$ & \small $9.55\pm0.19$ & \small $12.65\pm2.67$ & \small $12.42\pm0.33$ & \small $0.02\pm0.02$  \\
\hline
\end{tabular}
\\
\footnotesize{MAE = Mean Absolute Error in HR estimation (Beats/Min), RMSE = Root Mean Square Error in HR estimation (Beats/Min), MAPE = Mean Percentage Error (\%), Pearson = Pearson Correlation in HR estimation}
\end{table}

\subsection{Heart Rate (HR) Estimation}
\label{sec: heart_rate_estimation}

\subsubsection{Benchmark results on ring-based HR estimation}

In the HR estimation task, we evaluate the performance of different methods using the filtered (0.5-3~Hz) infrared channel PPG data from Ring 1 and Ring 2. 
Table~\ref{tab: hr_performance} presents performance results for HR estimation across all scenarios, with the best MAE of 5.18 BPM on Ring 1. Table~\ref{tab: hr_performance_stationary} and Table~\ref{tab: hr_performance_motion} show the performance of different methods in stationary and motion scenarios, respectively. In stationary scenarios, the best MAE is 3.71 BPM on Ring 1, while in motion scenarios, the best MAE is 16.10 BPM on Ring 2. Table~\ref{tab: hr_performance_commercial_samsung} and Table~\ref{tab: hr_performance_commercial_oura} compare our results with the built-in HR estimation methods of Samsung and Oura rings. The best MAE of our methods is 6.15 BPM on Ring 1, while the built-in method achieves 9.90 BPM on Samsung and 10.09 BPM on Oura.

\subsubsection{Insights and findings on ring-based HR estimation}

\vspace*{0.05in}\noindent\textbf{Our evaluated supervised learning methods outperform commercial ring devices.} The best-performing ResNet model achieves an MAE of 6.15 BPM on Ring 1, substantially better than commercial Samsung (9.90 BPM) and Oura (10.09 BPM) rings (Tables~\ref{tab: hr_performance_commercial_samsung} and \ref{tab: hr_performance_commercial_oura}). Additionally, these commercial devices may fail to provide measurements during motion, prompting users to remain still, which limits their utility in real-world applications.

\begin{table}[htbp]
    \centering
    \setlength{\tabcolsep}{3pt}  
    \caption{Benchmark results of different methods on HR task compared with Samsung Ring.}
    \label{tab: hr_performance_commercial_samsung}
    \begin{tabular}{r|cc|cc|cc|cc}
    \hline
    \textbf{HR} & \multicolumn{2}{c|}{\textbf{Ring 1 (Reflective)}} & \multicolumn{2}{c|}{\textbf{Samsung}} & \multicolumn{2}{c|}{\textbf{Ring 2 (Transmissive)}} & \multicolumn{2}{c}{\textbf{Samsung}}  \\
    \textbf{Comparison}&  \small MAE & \small MAPE & \small MAE & \small MAPE & \small MAE & \small MAPE & \small MAE & \small MAPE  \\
    \hline
    \hline
    \textbf{Physics-based} & & & \multicolumn{2}{c|}{\textbf{Built-in*}} & & &  \multicolumn{2}{c}{\textbf{Built-in*}}\\
    \small Peak & \small $18.93\pm1.43$ & \small $25.99\pm2.43$ & & & \small $17.18\pm1.56$ & \small $23.29\pm2.57$ & &\\
    \small FFT & \small $12.06\pm1.50$ & \small $13.72\pm1.57$ & & & \small $20.25\pm2.22$ & \small $22.30\pm2.20$ & & \\
    \textbf{Supervised} & & & & & & &  \\
    \small ResNet & \small \bm{$6.32\pm0.69$} & \small \bm{$8.72\pm1.30$} & \small{$9.90\pm1.25$}& \small{$13.94\pm2.34$} & \small \bm{$9.87\pm0.98$} & \small \bm{$12.30\pm1.54$} & \small{$9.94\pm1.57$}& \small{$13.32\pm2.83$}\\
    \small InceptionTime & \small $8.90\pm0.69$ & \small $11.65\pm1.21$ & & & \small \textbf{$10.22\pm0.89$} & \small $12.67\pm1.33$ & &  \\
    \small Transformer & \small $9.97\pm0.69$ & \small $12.78\pm1.29$ & & & \small $12.10\pm0.92$ & \small $14.95\pm1.36$ & & \\
    \small Mamba & \small $9.99\pm0.71$ & \small $13.72\pm1.43$ & & & \small $13.77\pm0.92$ & \small $17.44\pm1.54$ & &  \\
    \hline
    \end{tabular}
    \\
    \footnotesize{MAE = Mean Absolute Error in HR estimation (Beats/Min), Pearson = Pearson Correlation in HR estimation. Built-in*: The built-in HR estimation methods of Samsung Ring.}

\end{table}
\begin{table}[htbp]
    \centering
    \setlength{\tabcolsep}{3pt}  
    \caption{Benchmark results of different methods on HR task compared with Oura Ring.}
    \label{tab: hr_performance_commercial_oura}
    \begin{tabular}{r|cc|cc|cc|cc}
        \hline
        \textbf{HR} & \multicolumn{2}{c|}{\textbf{Ring 1 (Reflective)}} & \multicolumn{2}{c|}{\textbf{Oura}} & \multicolumn{2}{c|}{\textbf{Ring 2 (Transmissive)}} & \multicolumn{2}{c}{\textbf{Oura}}  \\
        \textbf{Comparison}&  \small MAE & \small MAPE & \small MAE & \small MAPE & \small MAE & \small MAPE & \small MAE & \small MAPE  \\
        \hline
        \hline
        \textbf{Physics-based} & & & \multicolumn{2}{c|}{\textbf{Built-in*}} & & &  \multicolumn{2}{c}{\textbf{Built-in*}}\\
        \small Peak & \small $19.82\pm1.32$ & \small $27.78\pm2.25$ & & & \small $15.75\pm1.14$ & \small $21.73\pm1.93$ & &\\
        \small FFT & \small $11.37\pm1.26$ & \small $13.16\pm1.34$ & & & \small $17.62\pm1.75$ & \small $20.06\pm1.79$ & & \\
        \textbf{Supervised} & & & & & & &  \\
        \small ResNet & \small \bm{$6.15\pm0.60$} & \small \bm{$8.73\pm1.16$} & \small{$10.09\pm1.13$}& \small{$14.67\pm2.23$} & \small \bm{$6.15\pm0.76$} & \small \bm{$10.57\pm1.20$} & \small{$10.56\pm1.43$}& \small{$14.38\pm2.63$}\\
        \small InceptionTime & \small $8.61\pm0.60$ & \small $11.58\pm1.06$ & & & \small \textbf{$9.44\pm0.72$} & \small $11.91\pm1.05$ & &  \\
        \small Transformer & \small $9.90\pm0.63$ & \small $12.97\pm1.17$ & & & \small $11.23\pm0.75$ & \small $14.26\pm1.11$ & & \\
        \small Mamba & \small $10.41\pm0.63$ & \small$14.57\pm1.27$ & & & \small $13.38\pm0.76$ & \small $17.47\pm1.27$ & &  \\
        \hline
    \end{tabular}
    \\
    \footnotesize{MAE = Mean Absolute Error in HR estimation (Beats/Min), Pearson = Pearson Correlation in HR estimation. Built-in*: The built-in HR estimation methods of Oura Ring.}
\end{table}

\vspace*{0.05in}\noindent\textbf{Supervised learning methods consistently outperform physics-based approaches across all scenarios, though motion conditions remain challenging.} The ResNet model achieves the best overall performance with an MAE of 5.18 BPM on Ring 1 across all scenarios, compared to 9.32 BPM for the best physics-based method (FFT). Moreover, ResNet demonstrates much stronger correlation with reference HR values (Pearson coefficient of 0.73 vs. 0.36 for FFT), indicating superior tracking of actual heart rate variations. During stationary conditions, the ResNet model achieves a best MAE of 3.71 BPM, with a strong correlation of 0.90 (Table~\ref{tab: hr_performance_stationary}). This performance is significantly better than the best physics-based method (6.04 BPM, Pearson coefficient of 0.56). 
Despite these improvements, even the best supervised models show substantial error increases during motion (from 3.71 BPM to 18.26 BPM), indicating a persistent gap between current capabilities and clinical requirements, which typically demand errors below 5 BPM for medical applications~\cite{morgadoareia2021chest}.

\begin{table}[htbp]
    \centering
    \caption{Benchmark results of different methods on HR task in stationary scenarios.}
    \setlength{\tabcolsep}{3pt}  
    \label{tab: hr_performance_stationary}
    \begin{tabular}{r|cccc|cccc}
    \hline
    \textbf{HR} & \multicolumn{4}{c}{\textbf{Ring 1 (Reflective)}} & \multicolumn{4}{c}{\textbf{Ring 2 (Transmissive)}} \\
    \textbf{Stationary}&  \small MAE & \small  RMSE &  
    \small  MAPE & \small  Pearson &  \small MAE & \small  RMSE &  
    \small  MAPE & \small  Pearson \\
    \hline
    \hline
    \textbf{Physics-based} & & & & & & & &  \\
    \small Peak & \small $18.17\pm0.46$ & \small $26.47\pm5.19$ & \small $23.49\pm0.60$ & \small $0.38\pm0.02$ & \small $28.50\pm0.55$ & \small $36.10\pm6.15$ & \small $37.93\pm0.76$ & \small $0.06\pm0.02$  \\
    \small FFT & \small $6.04\pm0.32$ & \small $14.50\pm3.97$ & \small $7.30\pm0.37$ & \small $0.56\pm0.02$ & \small$12.42\pm0.43$ & \small $21.40\pm4.77$ & \small $15.21\pm0.50$ & \small $0.18\pm0.02$  \\
    \textbf{Supervised} & & & & & & & &  \\
    \small ResNet & \small \bm{$3.71\pm0.09$} & \small \bm{$5.19\pm1.15$} & \small \bm{$4.77\pm0.12$} & \small \bm{$0.90\pm0.01$} & \small \bm{$7.15\pm0.16$} & \small \bm{$9.66\pm1.99$} & \small \bm{$9.16\pm0.20$} & \small \bm{$0.44\pm0.02$}  \\
    \small InceptionTime & \small $5.38\pm0.13$ & \small $7.53\pm1.68$ & \small $6.86\pm0.17$ & \small $0.77\pm0.02$ & \small $7.34\pm0.15$ & \small$9.54\pm1.89$ & \small \textbf{$9.49\pm0.20$} & \small $0.40\pm0.02$   \\
    \small Transformer & \small $6.80\pm0.14$ & \small $8.95\pm1.77$ & \small $8.35\pm0.17$ & \small $0.65\pm0.02$ & \small $7.82\pm0.15$ & \small $9.87\pm1.83$ & \small $10.15\pm0.19$ & \small $0.24\pm0.02$ \\
    \small Mamba & \small $6.45\pm0.14$ & \small \textbf{$8.70\pm2.18$} & \small$8.46\pm0.20$ & \small $0.71\pm0.02$ & \small $8.64\pm0.17$ & \small $11.00\pm2.04$ & \small $11.18\pm0.23$ & \small $0.05\pm0.02$  \\
    \hline
    \end{tabular}
    \\
    \footnotesize{MAE = Mean Absolute Error in HR estimation (Beats/Min), RMSE = Root Mean Square Error in HR estimation (Beats/Min), MAPE = Mean Percentage Error (\%), Pearson = Pearson Correlation in HR estimation}
    \end{table}

    \begin{table}[htbp]
        \centering
        \setlength{\tabcolsep}{3pt}  
        \caption{Benchmark results of different methods on HR task in motion scenarios.}
        \label{tab: hr_performance_motion}
        \begin{tabular}{r|cccc|cccc}
        \hline
        \textbf{HR} & \multicolumn{4}{c}{\textbf{Ring 1 (Reflective)}} & \multicolumn{4}{c}{\textbf{Ring 2 (Transmissive)}} \\
       \textbf{Motion} &  \small MAE & \small  RMSE &  
        \small  MAPE & \small  Pearson &  \small MAE & \small  RMSE &  
        \small  MAPE & \small  Pearson \\
        \hline
        \hline
        \textbf{Physics-based} & & & & & & & &  \\
        \small Peak & \small $21.27\pm1.42$ & \small$28.93\pm10.14$ & \small $35.86\pm3.84$ & \small $0.12\pm0.07$ & \small $28.85\pm1.52$ & \small $35.35\pm11.13$ & \small $41.07\pm3.60$ & \small $-0.26\pm0.07$  \\
        \small FFT & \small $39.25\pm1.63$ & \small $45.25\pm11.65$ & \small $43.65\pm1.37$ & \small $0.09\pm0.07$ & \small $39.52\pm1.69$ & \small $45.55\pm12.20$ & \small $42.79\pm1.41$ & \small $0.10\pm0.07$   \\
        \textbf{Supervised} & & & & & & & &  \\
        \small ResNet & \small $18.52\pm1.08$ & \small $23.79\pm7.33$ & \small$27.25\pm2.56$ & \small $-0.10\pm0.07$ & \small $16.74\pm1.08$ & \small $22.17\pm7.43$ & \small $22.65\pm2.38$& \small $-0.01\pm0.07$  \\
        \small InceptionTime & \small $18.71\pm1.06$ & \small $23.72\pm7.04$ & \small \bm{$26.36\pm2.25$} & \small $0.01\pm0.07$ & \small \bm{$16.10\pm0.98$} & \small \bm{$20.76\pm6.62$} & \small \bm{$21.23\pm2.06$} & \small \bm{$0.25\pm0.07$}   \\
        \small Transformer & \small \bm{$18.26\pm1.05$} & \small \bm{$23.28\pm7.33$} & \small $27.21\pm2.63$ & \small $0.04\pm0.07$ & \small $16.73\pm1.02$ & \small $21.59\pm6.82$ & \small $21.82\pm2.13$ & \small $0.06\pm0.07$ \\
    \small Mamba & \small $18.45 \pm 1.04$ & \small $23.34 \pm 7.15$ & \small $26.76 \pm 2.46$ & \small \bm{$0.06 \pm 0.07$} & \small $17.84 \pm 1.05$ & \small $22.74 \pm 7.40$ & \small $23.73 \pm 2.41$ & \small $-0.15 \pm 0.07$  \\
        \hline
        \end{tabular}
        \\
        \footnotesize{MAE = Mean Absolute Error in HR estimation (Beats/Min), RMSE = Root Mean Square Error in HR estimation (Beats/Min), MAPE = Mean Percentage Error (\%), Pearson = Pearson Correlation in HR estimation}
        \end{table}

\begin{figure}[htbp]
    \centering
    \includegraphics[width=0.9\columnwidth]{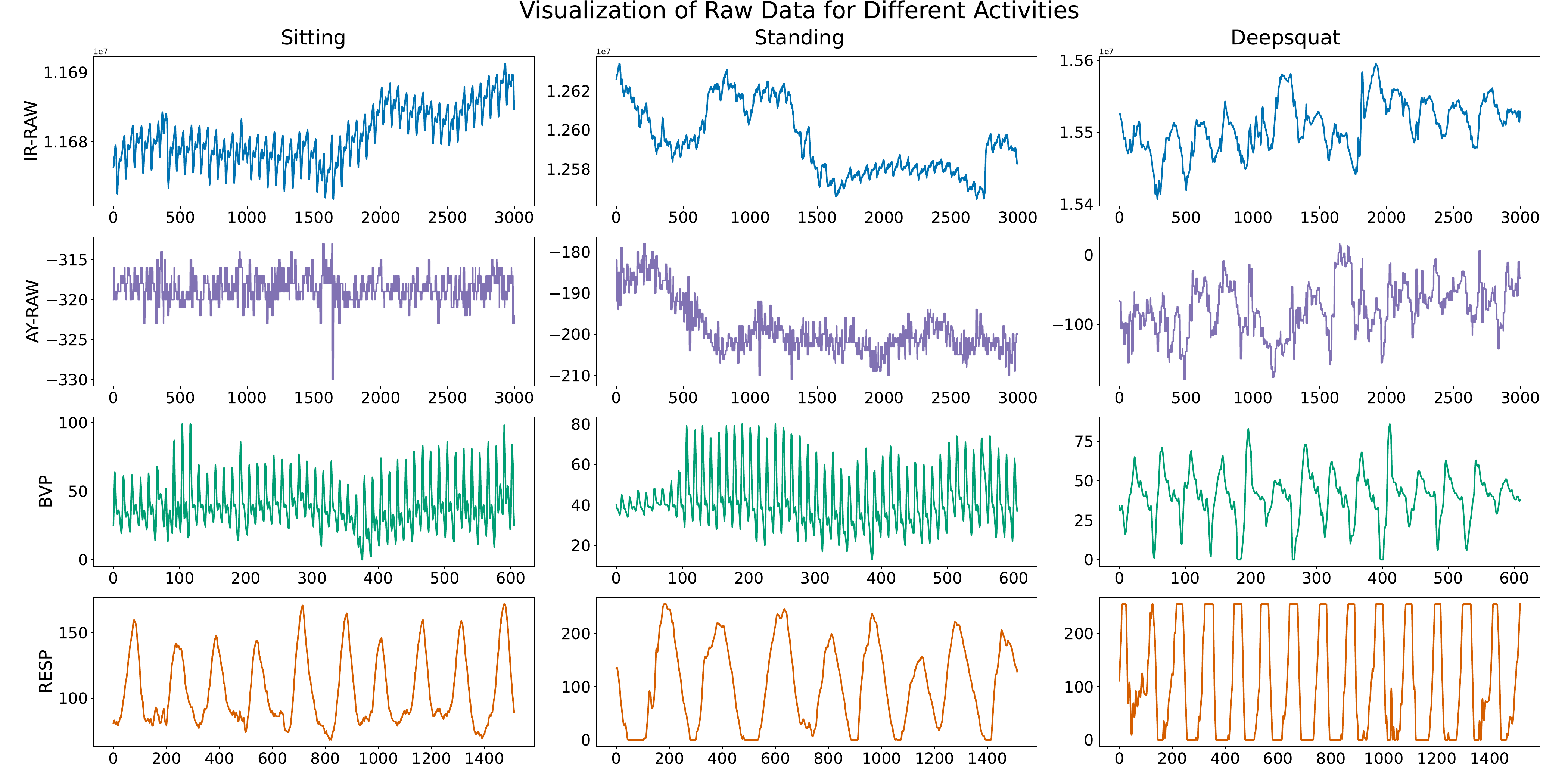}
    \caption{Visualization of ring signal and corresponding ground truth}
\label{fig: visualization}
\end{figure}

\vspace*{0.05in}\noindent\textbf{Ring 1 (reflective PPG) demonstrates superior performance in most scenarios, while Ring 2 (transmissive PPG) shows advantages during motion.} In stationary conditions, Ring 1 achieves an MAE of 3.71 BPM compared to Ring 2's 7.15 BPM (Table~\ref{tab: hr_performance_stationary}). However, during motion scenarios as Figure~\ref{fig: visualization} shows, Ring 2 delivers slightly better results with an MAE of 16.10 BPM versus Ring 1's 18.26 BPM (Table~\ref{tab: hr_performance_motion}). This performance difference during motion can be attributed to the higher light intensity used in transmissive PPG sensing. This greater illumination power creates a stronger signal-to-noise ratio when motion artifacts occur, as the physiological signal component maintains a relatively higher amplitude compared to noise. Consequently, while Ring 2's transmissive approach may underperform in stationary scenarios due to placement and configuration limitations, its inherent advantage in signal strength proves potential against motion artifacts.

\vspace*{0.05in}\noindent\textbf{Physics-based methods show acceptable performance in stationary scenarios but degrade substantially under motion conditions.} The FFT approach achieves a reasonable 6.04 BPM MAE with Ring 1 during stationary periods (Table ~\ref{tab: hr_performance_stationary}), but its error increases dramatically to 39.25 BPM during motion (Table ~\ref{tab: hr_performance_motion}). This illustrates how traditional signal processing techniques lack robustness against motion artifacts, making them unsuitable for real-world applications where user movement is expected.

\begin{table}[htbp]
    \centering
    \setlength{\tabcolsep}{3pt}
    \caption{Ablation study of multi-channel fusion on HR estimation in motion scenarios with Mamba model}
    \label{tab: ablation_study}
    \begin{tabular}{r|cccc|cccc}
    \hline
    \textbf{HR} & \multicolumn{4}{c}{\textbf{Ring 1 (Reflective)}} & \multicolumn{4}{c}{\textbf{Ring 2 (Transmissive)}} \\
   \textbf{Motion} &  \small MAE & \small  RMSE &  
    \small  MAPE & \small  Pearson &  \small MAE & \small  RMSE &  
    \small  MAPE & \small  Pearson \\
    \hline
    \hline
    \textbf{In-Channel(s)}  \\
    \small IR & \small $18.45 \pm 1.04$ & \small $23.34 \pm 7.15$ & \small $26.76 \pm 2.46$ & \small $0.06 \pm 0.07$ & \small $17.84 \pm 1.05$ & \small $22.74 \pm 7.40$ & \small $23.73 \pm 2.41$ & \small $-0.15 \pm 0.07$  \\
    % \small RED & \small $16.95 \pm 1.03$ & \small $22.11 \pm 7.08$ & \small $25.84 \pm 2.59$ & \small $0.23 \pm 0.07$ & \small $16.56 \pm 1.05$ & \small $21.73 \pm 7.30$ & \small $22.50 \pm 2.43$ & \small $-0.19 \pm 0.07$  \\
    % \small IR+RED & \small $19.14 \pm 1.08$ & \small $24.22 \pm 7.42$ & \small $27.87 \pm 2.50$ & \small $-0.04 \pm 0.07$ & \small $16.64 \pm 1.00$ & \small $21.38 \pm 6.74$ & \small $21.63 \pm 2.02$ & \small $0.14 \pm 0.07$  \\
    \small IR+ACC & \small $17.63 \pm 1.04$ & \small $22.68 \pm 6.97$ & \small $26.04 \pm 2.48$ & \small $0.10 \pm 0.07$ & \small $16.58 \pm 1.07$ & \small $21.93 \pm 7.17$ & \small $22.30 \pm 2.30$ & \small $-0.13 \pm 0.07$  \\
    % \small RED+ACC & \small $17.89 \pm 1.04$ & \small $22.92 \pm 7.15$ & \small $26.63 \pm 2.58$ & \small $-0.00 \pm 0.07$ & \small $16.42 \pm 1.05$ & \small $21.66 \pm 7.08$ & \small $21.97 \pm 2.25$ & \small $-0.16 \pm 0.07$  \\    
\small IR+RED+ACC & \small \bm{$17.11 \pm 1.05$} & \small \bm{$22.46 \pm 7.13$} & \small \bm{$25.97 \pm 2.60$} & \small \bm{$0.15 \pm 0.07$} & \small \bm{$15.21 \pm 1.01$} & \small \bm{$20.38 \pm 6.87$} & \small \bm{$21.07 \pm 2.35$} & \small \bm{$0.05 \pm 0.07$}  \\

    \hline
    \end{tabular}
    \\
    \footnotesize{MAE = Mean Absolute Error in HR estimation (Beats/Min), RMSE = Root Mean Square Error in HR estimation (Beats/Min), MAPE = Mean Percentage Error (\%), Pearson = Pearson Correlation in HR estimation}
\end{table}

\vspace*{0.05in}\noindent\textbf{Multi-channel fusion significantly improves HR estimation accuracy in motion scenarios.} In an ablation study comparing different input combinations (IR, IR+ACC, IR+RED+ACC), the fusion of IR, RED, and ACC consistently achieved the best results (Table~\ref{tab: ablation_study}). For Ring 2, the MAE improved from $17.84 \pm 1.05$ BPM using IR alone to $15.21 \pm 1.01$ BPM with full fusion. Similarly, for Ring 1, MAE decreased from $18.45 \pm 1.04$ BPM (IR only) to $17.11 \pm 1.05$ BPM. These improvements highlight the importance of ACC signals in mitigating motion artifacts by capturing dynamic variations, while RED contributes additional physiological information. Overall, multi-channel fusion—especially with ACC integration—proves effective in reducing motion-induced distortions and enhancing the robustness of cardiovascular sensing in motion scenarios.

\subsection{Respiratory Rate (RR) Estimation}
\label{sec: respiratory_rate_estimation}

\subsubsection{Benchmark results on ring-based RR estimation}
In the RR estimation task, we evaluate the performance of different methods using the filtered (0.1- 0.5~Hz) infrared channel PPG data from Ring 1 and Ring 2. 
Table~\ref{tab: rr_performance} presents performance results for RR estimation across all scenarios, with the best MAE of 2.98 breaths per minute (BPM) achieved by the peak detection method on Ring 1. Table~\ref{tab: rr_performance_stationary} and Table~\ref{tab: rr_performance_motion} show the performance of different methods in stationary and motion scenarios, respectively. In stationary scenarios, the best MAE is 2.85 BPM on Ring 1, while in motion scenarios, the best MAE is 4.15 BPM on Ring 1.

\begin{table}[htbp]
    \centering
    \setlength{\tabcolsep}{3pt}  
    \caption{Benchmark results of different methods on RR estimation.}
    \label{tab: rr_performance}
    \begin{tabular}{r|cccc|cccc}
        \hline
        \textbf{RR} & \multicolumn{4}{c}{\textbf{Ring 1 (Reflective)}} & \multicolumn{4}{c}{\textbf{Ring 2 (Transmissive)}} \\
        \textbf{All} &  \small MAE & \small  RMSE &  
        \small  MAPE & \small  Pearson &  \small MAE & \small  RMSE &  
        \small  MAPE & \small  Pearson \\
        \hline
        \hline
        \textbf{Physics-based} & & & & & & & &  \\
        \small Peak & \small \bm{$2.98\pm0.07$} & \small $4.12\pm0.85$ & \small \bm{$18.39\pm0.50$} & \small \bm{$0.50\pm0.02$} & \small $3.84\pm0.08$ & \small $5.19\pm1.04$ & \small $24.62\pm0.72$& \small \bm{$0.22\pm0.02$}  \\
        \small FFT & \small $4.83\pm0.08$ & \small $5.97\pm1.09$ & \small $26.35\pm0.37$ & \small $0.33\pm0.02$ & \small $4.96\pm0.11$ & \small $6.01\pm1.24$ & \small $27.08\pm0.56$ & \small $0.21\pm0.03$   \\
        \textbf{Supervised} & & & & & & & &  \\
        \small ResNet & \small $3.16\pm0.06$ & \small \bm{$4.10\pm0.80$} & \small $20.45\pm0.52$ & \small$0.29\pm0.02$ & \small $3.34\pm0.07$ & \small $4.30\pm0.83$ & \small$20.70\pm0.57$ & \small $0.18\pm0.02$  \\
        \small InceptionTime & \small $3.25\pm0.06$ & \small$4.20\pm0.82$ & \small $20.78\pm0.52$ & \small $0.23\pm0.02$ & \small $3.36\pm0.07$ & \small $4.33\pm0.83$ & \small \bm{$20.70\pm0.56$} & \small $0.16\pm0.02$   \\
        \small Transformer & \small$3.35\pm0.06$ & \small $4.36\pm0.84$ & \small $21.39\pm0.55$ & \small $-0.11\pm0.02$ & \small $3.33\pm0.07$ & \small $4.32\pm0.84$ & \small $20.92\pm0.59$ & \small $0.08\pm0.02$ \\
        \small Mamba & \small $3.29\pm0.06$ & \small \textbf{$4.23\pm0.79$} & \small $22.20\pm0.60$ & \small $0.15\pm0.02$ & \small \bm{$3.30\pm0.07$} & \small\bm{$4.28\pm0.83$} & \small $21.20\pm0.62$ & \small $0.13\pm0.02$  \\
        \hline
        \end{tabular}
        \\
    \footnotesize{MAE = Mean Absolute Error in RR estimation (Breaths/Min), RMSE = Root Mean Square Error in RR estimation (Breaths/Min), MAPE = Mean Percentage Error (\%), Pearson = Pearson Correlation in RR estimation}
\end{table}

\subsubsection{Insights and findings on ring-based RR estimation}

\vspace*{0.05in}\noindent\textbf{Ring-based respiratory rate monitoring demonstrates promising accuracy, particularly in stationary conditions.} To the best of our knowledge, our benchmark is the first evaluation of respiratory rate estimation on rings. The best methods achieve an MAE of 2.98 BPM on Ring 1 and 3.30 BPM on Ring 2 (Table~\ref{tab: rr_performance}). This performance approaches clinical requirements of errors below 3 BPM~\cite{morgadoareia2021chest}. PPG sensors can capture respiratory information due to the cardiorespiratory coupling phenomenon~\cite{dick2014cardiorespiratory}, where breathing causes variations in cardiac output, blood pressure, and peripheral blood volume that modulate the PPG waveform's baseline, amplitude, and frequency characteristics. While performance remains stable in stationary scenarios with an MAE of 2.85 BPM on Ring 1, accuracy decreases during movement (4.15 BPM), indicating that motion robustness requires further improvement. This pattern suggests ring-based PPG sensors show potential for reliable RR monitoring in resting conditions, though their performance under motion still needs validation for real-world applications.

\vspace*{0.05in}\noindent\textbf{For Ring 1 (reflective), traditional physics-based peak detection methods outperform both FFT and supervised learning approaches.} The peak detection method achieves the lowest error (MAE of 2.98 BPM) and highest correlation with ground truth (Pearson coefficient of 0.50) across all scenarios (Table~\ref{tab: rr_performance}). This superior performance likely stems from the rhythmic nature of breathing patterns, which create distinctive amplitude modulations in the PPG signal that peak detection algorithms can effectively isolate. In contrast, FFT methods show substantially higher errors (4.83 BPM), possibly due to their sensitivity to non-respiratory frequency components in the signal.

\vspace*{0.05in}\noindent\textbf{For Ring 2 (transmissive), neural network-based approaches demonstrate better overall performance, with different architectures excelling in different scenarios.} In stationary conditions, the Mamba model achieves the best results with an MAE of 3.08 BPM (Table~\ref{tab: rr_performance_stationary}), likely due to its effective sequential modeling capabilities for regular breathing patterns. However, during motion, the ResNet model performs best with an MAE of 4.70 BPM (Table~\ref{tab: rr_performance_motion}), suggesting its convolutional architecture may better extract respiratory features from noisy motion-affected signals. This pattern indicates that transmissive PPG signals may contain more complex respiratory information that benefits from deep learning's feature extraction capabilities.

\begin{table}[htbp]
    \centering
    \setlength{\tabcolsep}{3pt}  
    \caption{Benchmark results of different methods on RR estimation in stationary scenarios.}
    \label{tab: rr_performance_stationary}
    \begin{tabular}{r|cccc|cccc}
        \hline
        \textbf{RR} & \multicolumn{4}{c}{\textbf{Ring 1 (Reflective)}} & \multicolumn{4}{c}{\textbf{Ring 2 (Transmissive)}} \\
        \textbf{Stationary}  &\small MAE & \small  RMSE &  
        \small  MAPE & \small  Pearson &  \small MAE & \small  RMSE &  
        \small  MAPE & \small  Pearson \\
        \hline
        \hline
        \textbf{Physics-based} & & & & & & & &  \\
        \small Peak & \small \bm{$2.85\pm0.07$} & \small $3.94\pm0.85$ & \small \bm{$18.22\pm0.53$} & \small \bm{$0.44\pm0.02$} & \small $3.72\pm0.09$ & \small $5.06\pm1.06$ & \small $24.32\pm0.75$& \small $0.19\pm0.03$  \\
        \small FFT & \small $4.58\pm0.08$ & \small $5.56\pm1.07$ & \small $26.08\pm0.40$ & \small $0.27\pm0.03$ & \small $4.93\pm0.11$ & \small $5.98\pm1.24$ & \small $26.95\pm0.55$ & \small \bm{$0.21\pm0.03$}   \\
        \textbf{Supervised} & & & & & & & &  \\
        \small ResNet & \small $2.98\pm0.06 $ & \small \bm{$3.81\pm0.76$} & \small $20.47\pm0.59$ & \small$0.20\pm0.02$ & \small $3.35\pm0.07$ & \small $4.23\pm0.84$ & \small$21.31\pm0.59$ & \small $-0.03\pm0.03$  \\
        \small InceptionTime & \small $2.96\pm0.06$ & \small$3.86\pm0.78$ & \small $20.44\pm0.61$ & \small $0.08\pm0.02$ & \small $3.19\pm0.07$ & \small $4.09\pm0.82$ & \small \bm{$20.29\pm0.58$} & \small $0.03\pm0.03$   \\
        \small Transformer & \small $3.11\pm0.06$ & \small $4.00\pm0.78$ & \small $21.30\pm0.60$ & \small $-0.13\pm0.02$ & \small $3.18\pm0.07$ & \small $4.09\pm0.83$ & \small $20.53\pm0.60$ & \small $-0.06\pm0.03$ \\
        \small Mamba & \small $3.11\pm0.06$ & \small \textbf{$4.00\pm0.77$} & \small $22.21\pm0.65$ & \small $-0.00\pm0.02$ & \small \bm{$3.08\pm0.06$} & \small \bm{$3.99\pm0.81$} & \small $20.66\pm0.65$ & \small $0.14\pm0.03$ \\
        \hline
        \end{tabular}
        \\
    \footnotesize{MAE = Mean Absolute Error in RR estimation (Breaths/Min), RMSE = Root Mean Square Error in RR estimation (Breaths/Min), MAPE = Mean Percentage Error (\%), Pearson = Pearson Correlation in RR estimation}
\end{table}

\begin{table}[htbp]
    \centering
    \setlength{\tabcolsep}{3pt}  
    \caption{Benchmark results of different methods on RR estimation in motion scenarios.}
    \label{tab: rr_performance_motion}
    \begin{tabular}{r|cccc|cccc}
        \hline
        \textbf{RR} & \multicolumn{4}{c}{\textbf{Ring 1 (Reflective)}} & \multicolumn{4}{c}{\textbf{Ring 2 (Transmissive)}} \\
        \textbf{Motion}&  \small MAE & \small  RMSE &  
        \small  MAPE & \small  Pearson &  \small MAE & \small  RMSE &  
        \small  MAPE & \small  Pearson \\
        \hline
        \hline
        \textbf{Physics-based} & & & & & & & &  \\
        \small Peak & \small \bm{$4.15\pm0.25$} & \small \bm{$5.41\pm1.73$} & \small \bm{$19.88\pm1.50$} & \small \bm{$0.30\pm0.0$} & \small $4.90\pm0.29$ & \small $6.23\pm2.02$ & \small $27.24\pm2.56$& \small $0.14\pm0.08$  \\
        \small FFT & \small $6.54\pm0.73$ & \small $7.64\pm3.41$ & \small $33.00\pm3.2$ & \small $0.04\pm0.19$ & \small $6.45\pm1.06$ & \small $7.58\pm4.10$ & \small $35.22\pm6.75$ & \small $0.20\pm0.27$   \\
        \textbf{Supervised} & & & & & & & &  \\
              \small ResNet & \small $5.00\pm0.25$ & \small $6.09\pm1.75$ & \small $22.67\pm1.20$ & \small$0.01\pm0.07$ & \small \bm{$4.70\pm0.27$} & \small \bm{$5.92\pm1.79$} & \small \bm{$23.73\pm2.02$} & \small \bm{$0.20\pm0.08$}  \\
        \small InceptionTime & \small $5.14\pm0.25$ & \small$6.17\pm1.72$ & \small $23.27\pm1.19$ & \small $0.01\pm0.07$ & \small $4.87\pm0.28$ & \small $6.08\pm1.80$ & \small $24.32\pm1.99$ & \small $0.19\pm0.08$   \\
        \small Transformer & \small $5.47\pm0.25$ & \small$6.50\pm1.82$ & \small$24.34\pm1.17$ & \small $-0.02\pm0.07$ & \small $5.31\pm0.28$ & \small $6.42\pm1.82$ & \small $26.02\pm1.95$ & \small $0.12\pm0.08$ \\
        \small Mamba & \small $4.88\pm0.23$ & \small \textbf{$5.85\pm1.67$} & \small $22.15\pm1.28$ & \small $0.19\pm0.07$ & \small $5.21\pm0.27$ & \small $6.31\pm1.80$ & \small $26.04\pm2.13$ & \small $0.04\pm0.08$  \\
        \hline
        \end{tabular}
        \\
    \footnotesize{MAE = Mean Absolute Error in RR estimation (Breaths/Min), RMSE = Root Mean Square Error in RR estimation (Breaths/Min), MAPE = Mean Percentage Error (\%), Pearson = Pearson Correlation in RR estimation}
\end{table}

\subsection{Blood Oxygen Saturation (SpO2) Estimation}
\label{sec: blood_oxygen_saturation_estimation}

\subsubsection{Benchmark results on ring-based SpO2 estimation}

Table~\ref{tab: spo2_performance} presents the performance evaluation of SpO2 estimation methods using only data from low oxygen simulation scenarios, as normal SpO2 levels (above 95\%) provide insufficient variation for meaningful assessment. For physics-based ratio method, we use the raw PPG signal from the infrared and red channels of Ring 1 and Ring 2. For supervised learning methods, we use the filtered (0.5–3~Hz) PPG signals instead of raw signals due to the large values in raw data, which hinder model convergence. For Ring 1 (Reflective), the ResNet model achieves the best performance with an MAE of 3.22\%. For Ring 2 (Transmissive), the physics-based Ratio method delivers the best results with an MAE of 3.31\% and a Pearson correlation coefficient of 0.28.

\subsubsection{Insights and findings on ring-based SpO2 estimation}

\vspace*{0.05in}\noindent\textbf{Physics-based Ratio method demonstrates stronger correlation with ground truth.} The Ratio method achieves higher Pearson correlation coefficients (0.30 for Ring 1 and 0.28 for Ring 2) compared to supervised learning approaches that show near-zero correlations. However, this method requires calibration of a and b parameters for each specific ring and potentially each user, limiting its scalability for widespread deployment.

\vspace*{0.05in}\noindent\textbf{Supervised learning methods achieve competitive error metrics but lack trend capturing ability.} Deep learning models like ResNet achieve comparable or slightly better MAE values (3.22\% on Ring 1) but their low correlation coefficients suggest they may be learning the data distribution rather than capturing the true physiological trends in blood oxygen levels.

\vspace*{0.05in}\noindent\textbf{Different ring designs show varying affinities for estimation methods.} Ring 1 works better with deep learning approaches (lowest MAE of 3.22\%), while Ring 2 shows superior performance with the physics-based Ratio method (MAE of 3.31\%). This difference likely stems from the transmissive sensor configuration in Ring 2 providing cleaner signals for the ratio-based calculation, which aligns with traditional pulse oximetry principles.

\vspace*{0.05in}\noindent\textbf{Current ring-based SpO2 estimation faces fundamental limitations.} The similar error ranges across all methods (approximately 3-4\% MAE) indicate that advances in both sensor hardware and algorithm design are needed to improve accuracy for clinical applications.

\begin{table}[htbp]
    \centering
    \setlength{\tabcolsep}{3pt}  
    \caption{Benchmark results of different methods on SpO2 estimation in low oxygen simluation scenario.}
    \label{tab: spo2_performance}
    \begin{tabular}{r|cccc|cccc}
        \hline
        \textbf{SpO2} & \multicolumn{4}{c}{\textbf{Ring 1 (Reflective)}} & \multicolumn{4}{c}{\textbf{Ring 2 (Transmissive)}} \\
        \textbf{Low Oxygen}&  \small MAE & \small  RMSE &  
        \small  MAPE & \small  Pearson &  \small MAE & \small  RMSE &  
        \small  MAPE & \small  Pearson \\
        \hline
        \hline
        \textbf{Physics-based} & & & & & & & &  \\
        \small Ratio & \small $3.57\pm0.14$ & \small $4.65\pm1.34$ & \small $3.86\pm0.16$ & \small\bm{$0.30\pm0.04$} & \small \bm{$3.31\pm0.13$} & \small \bm{$3.95\pm1.15$} & \small \bm{$3.58\pm0.15$} & \small \bm{$0.28\pm0.06$}   \\
        \textbf{Supervised} & & & & & & & &  \\
        \small ResNet & \small \bm{$3.22\pm0.14$} & \small \bm{$4.46\pm1.35$} & \small \bm{$3.57\pm0.17$} & \small $0.01\pm0.05$ & \small $3.62\pm0.22$ & \small $5.02\pm1.61$ & \small $4.03\pm0.25$ & \small $0.11\pm0.06$  \\
        \small InceptionTime & \small $3.31\pm0.15$ & \small $4.57\pm1.37$ & \small $3.66\pm0.18$ & \small $-0.01\pm0.05$ & \small $3.55\pm0.22$ & \small $4.97\pm1.60$ & \small $3.96\pm0.25$ & \small $0.03\pm0.06$   \\
        \small Transformer & \small $3.23\pm0.15$ & \small $4.53\pm1.37$ & \small $3.59\pm0.18$ & \small $0.01\pm0.05 $ & \small $3.56\pm0.22$ & \small $5.01\pm1.63$ & \small $3.97\pm0.26$ & \small $0.07\pm0.06$  \\
        \small Mamba & \small $3.62\pm0.17$ & \small $5.12\pm1.51$ & \small $4.04\pm0.20$ & \small $-0.08\pm0.05$ & \small $3.37\pm0.19$ & \small $4.52\pm1.46$ & \small $3.71\pm0.22$ & \small $0.13\pm0.06$  \\
        \hline
    \end{tabular}
    \\
    \footnotesize{MAE = Mean Absolute Error in SpO2 estimation, RMSE = Root Mean Square Error in SpO2 estimation, MAPE = Mean Percentage Error (\%), Pearson = Pearson Correlation in SpO2 estimation}
\end{table}

\subsection{Blood Pressure (BP) Estimation}
\label{sec: blood_pressure_estimation}

\subsubsection{Benchmark results on ring-based BP estimation}

In the BP estimation task, we only evaluate the performance of different supervised learning methods using the filtered (0.5-3~Hz) infrared and red PPG from Ring 1 and Ring 2 as there are few widely-used physics-based methods. Table ~\ref{tab: sbp_performance} and Table ~\ref{tab: dbp_performance} summarize the performance results of SBP and DBP estimation. For SBP estimation, the Transformer model performs best on Ring 1 with an MAE of 13.33 mmHg, while the Mamba architecture shows superior performance on Ring 2 with an MAE of 14.56 mmHg. For DBP estimation, the Transformer model achieves the best performance with an MAE of 7.56 mmHg on Ring 1 and 8.06 mmHg on Ring 2. 

\subsubsection{Insights and findings on ring-based BP estimation}

\vspace*{0.05in}\noindent\textbf{Transformer architecture demonstrates superior performance for BP estimation across both ring types.} The Transformer model achieves the lowest errors for both SBP (13.33 mmHg on Ring 1) and DBP (7.56 mmHg on Ring 1, 8.06 mmHg on Ring 2), suggesting its attention mechanism effectively captures the temporal relationships in PPG signals that correlate with blood pressure values. This advantage is particularly evident in the higher correlation coefficient (0.25) achieved for SBP estimation on Ring 1.

\vspace*{0.053in}\noindent\textbf{Ring 1 (reflective) consistently outperforms Ring 2 (transmissive) for BP estimation.} Across all models and metrics, the reflective PPG approach shows better results than the transmissive approach. This performance difference suggests that reflective sensors may better capture the pulsatile blood flow patterns that correlate with blood pressure, with less interference from surrounding tissues.

\vspace*{0.05in}\noindent\textbf{SBP estimation presents greater challenges than DBP estimation.} Error metrics for SBP (MAE of 13-17 mmHg) are consistently higher than those for DBP (MAE of 7-11 mmHg), indicating that systolic pressure is inherently more difficult to estimate non-invasively from PPG signals. This pattern aligns with previous findings in the field of cuffless BP monitoring~\cite{gonzalez2023benchmark}.

\vspace*{0.05in}\noindent\textbf{Current ring-based BP estimation approaches fall short of clinical standards.} With MAPE values around 11-15\% for both SBP and DBP and generally low correlation coefficients, our results highlight the gap between current ring-based methods and clinical requirements (typically ±5 mmHg). This performance gap underscores the need for improved sensor technology and more advanced algorithms for wearable BP monitoring.

\begin{table}[htbp]
    \centering
    \setlength{\tabcolsep}{3pt}  
    \caption{Benchmark results of different methods on SBP estimation.}
    \label{tab: sbp_performance}
    \begin{tabular}{r|cccc|cccc}
        \hline
        \textbf{SBP} & \multicolumn{4}{c}{\textbf{Ring 1 (Reflective)}} & \multicolumn{4}{c}{\textbf{Ring 2 (Transmissive)}} \\
       \textbf{All} &  \small MAE & \small  RMSE &  
        \small  MAPE & \small  Pearson &  \small MAE & \small  RMSE &  
        \small  MAPE & \small  Pearson \\
        \hline
        \hline
        \textbf{Supervised} & & & & & & & &  \\
        \small ResNet & \small $14.52\pm0.97$ & \small $18.38\pm6.34$ & \small $13.05\pm0.77$ & \small $0.18\pm0.09$ & \small $17.47\pm1.15$ & \small $20.78\pm6.90$ & \small $16.75\pm1.12$ & \small $-0.06\pm0.10$  \\
        \small InceptionTime & \small $14.79\pm0.93$ & \small $18.29\pm6.24$ & \small $13.54\pm0.77$ & \small $0.08\pm0.09$ & \small $17.03\pm1.15$ & \small $20.41\pm7.24$ & \small $15.53\pm0.92$ & \small $-0.07\pm0.10$   \\
        \small Transformer & \small \bm{$13.33\pm0.78$} & \small \bm{$16.10\pm5.02$} & \small \bm{$12.53\pm0.73$} & \small \bm{$0.25\pm0.08$} & \small $15.16\pm1.01$ & \small $18.08\pm6.26$ & \small $14.44\pm0.99$ & \small $0.03\pm0.10$ \\
        \small Mamba & \small $13.99\pm0.88$ & \small $17.30\pm5.75$ & \small $13.12\pm0.79$ & \small $-0.02\pm0.09$ & \small \bm{$14.56\pm1.04$} & \small \bm{$17.76\pm6.28$} & \small \bm{$13.79\pm0.98$} & \small \bm{$0.09\pm0.10$}  \\
        \hline
    \end{tabular}
    \\
    \footnotesize{MAE = Mean Absolute Error in SBP estimation (mmHg), RMSE = Root Mean Square Error in SBP estimation (mmHg), MAPE = Mean Percentage Error (\%), Pearson = Pearson Correlation in SBP estimation}
\end{table}

\begin{table}[htbp]
    \centering
    \setlength{\tabcolsep}{3pt}  
    \caption{Benchmark results of different methods on DBP  estimation.}
    \label{tab: dbp_performance}
    \begin{tabular}{r|cccc|cccc}
    \hline
    \textbf{DBP} & \multicolumn{4}{c}{\textbf{Ring 1 (Reflective)}} & \multicolumn{4}{c}{\textbf{Ring 2 (Transmissive)}} \\
 \textbf{All} &  \small MAE & \small  RMSE &  
    \small  MAPE & \small  Pearson &  \small MAE & \small  RMSE &  
    \small  MAPE & \small  Pearson \\
    \hline
    \hline
    \textbf{Supervised} & & & & & & & &  \\
    \small ResNet & \small $8.48\pm0.53$ & \small $10.51\pm3.56$ & \small $13.10\pm0.86$ & \small $-0.14\pm0.09$ & \small $9.73\pm0.78$ & \small $12.39\pm4.85$ & \small $15.61\pm1.44$ & \small $-0.14\pm0.10$  \\
    \small InceptionTime & \small $8.68\pm0.60$ & \small $11.16\pm4.15$ & \small $13.12\pm0.93$ & \small \bm{$0.01\pm0.09$} & \small $11.16\pm0.70$ & \small $13.10\pm4.48$ & \small $16.49\pm0.90$ & \small \bm{$-0.05\pm0.10$}   \\
    \small Transformer & \small \bm{$7.56\pm0.54$} & \small \bm{$9.80\pm3.49$} & \small \bm{$11.80\pm0.91$} & \small $-0.02\pm0.09$ & \small \bm{$8.06\pm0.69$} & \small \bm{$10.54\pm4.16$} & \small \bm{$12.57\pm1.14$} & \small $-0.15\pm0.10$ \\
    \small Mamba & \small $9.67\pm1.22$ & \small $17.19\pm12.62$ & \small $15.20\pm2.01$ & \small $-0.10\pm0.09$ & \small $9.94\pm0.79$ & \small $12.59\pm4.66$ & \small $15.94\pm1.41$ & \small $-0.07\pm0.10$  \\
    \hline
    \end{tabular}
    \\
    \footnotesize{MAE = Mean Absolute Error in DBP estimation (mmHg), RMSE = Root Mean Square Error in DBP estimation (mmHg), MAPE = Mean Percentage Error (\%), Pearson = Pearson Correlation in DBP estimation}
\end{table}

\section{Discussion}
\label{sec: discussion}

In this section, we discuss how our $\tau$-Ring dataset and RingTool benchmark contribute to reliable ring-based cardiovascular physiology sensing in Section~\ref{sec: discussion sub1}. We also analyze the advantages of ring-type devices for cardiovascular physiology sensing in Section~\ref{sec: discussion sub2}. Finally, we address the limitations of our dataset and benchmark, as well as potential future work in Section~\ref{sec: discussion sub3}.

\subsection{Reliable Ring-based Cardiovascular Physiology Sensing for Multiple Vital Signs}
\label{sec: discussion sub1}

The $\tau$-Ring dataset represents the first comprehensive ring-based health dataset that captures cardiovascular physiology across multiple scenarios. Our dataset offers several advantages that make it suitable for real-world applications. First, it features a wide distribution of physiological labels across four vital signs (HR, RR, SpO2, and BP) as Table~\ref{tab:activitycollection} shows, enabling algorithm development for multiple cardiovascular parameters simultaneously. Second, it includes two sensor configurations (Figure ~\ref{fig: ring_demonstrate}) that reflect common hardware setups in commercial ring devices, allowing for comparative studies. Third, the dataset spans diverse scenarios including both resting states and physical activities (Figure ~\ref{fig: daily_experiment}), as well as normal and abnormal physiological conditions (Figure ~\ref{fig: health_experiment}), which mirrors the complexity of real-world deployment environments.

By open-sourcing both our dataset and toolkit code at \datalink and \githublink, we aim to establish reliability and reproducibility standards in ring-based physiological sensing research. Our initial benchmarking results demonstrate that for resting scenarios, HR and RR measurements approach clinical accuracy standards, with average errors of approximately 5 beats per minute and 3 breaths per minute respectively. Compared to commercial rings like Samsung and Oura, our ring-based system shows even better performance in HR estimation (Ours: 6.15 BPM, Samsung: 9.90 BPM, Oura: 10.09 BPM), demonstrating the potential of ring devices for accurate cardiovascular monitoring. 
However, SpO2 and BP measurements, along with sensing during motion, present ongoing challenges with higher error rates (as detailed in Table~\ref{tab: all_performance}). These findings highlight the current capabilities and limitations of ring-based sensing technologies.
We encourage researchers to build upon this foundation by developing algorithms with improved accuracy and enhanced robustness across different scenarios. 

\subsection{Advantages of ring-type device for Cardiovascular Physiology Sensing}
\label{sec: discussion sub2}

Ring-based devices offer several distinct advantages for cardiovascular physiology sensing. The finger location provides thinner skin compared to the wrist, enabling PPG sensors to capture higher quality photoplethysmographic signals. Furthermore, rings afford continuous, 24/7 monitoring capabilities, as they can be comfortably worn during daily activities and sleep without the obtrusiveness that affects user compliance with other form factors such as earbuds.

Our benchmarking reveals that existing algorithms have achieved promising performance on ring-based data. For instance, ResNet-based approaches achieved mean absolute errors of approximately 5 BPM for HR estimation, while Peak detection methods demonstrated strong performance in RR estimation with errors around 3 breaths per minute. For more complex parameters like BP, Transformer-based architectures showed potential but require further refinement. However, these implementations are not necessarily optimal for ring form factors. Our preliminary ablation studies suggest that Mamba-based methods with ACC fusion can reduce motion artifacts and improve accuracy. Future work should explore multimodal fusion approaches that leverage multiple sensor channels to develop more robust algorithms capable of maintaining performance across varied scenarios, from rest to vigorous physical activity.

\subsection{Limitations and Future Work}
\label{sec: discussion sub3}

\subsubsection{Limits of $\tau$-Ring Dataset}
\label{sec: discussion sub3.1}
While $\tau$-Ring represents the first ring-based physiological dataset, it has several limitations. The dataset's scale could be expanded to include longer periods from more participants with greater demographic diversity. The blood pressure labels could benefit from more frequent sampling to better capture temporal variations. Regarding modalities, we omitted certain data types that could enhance sensing capabilities - specifically, green channel PPG data (to maintain consistency between reflective and transmissive modes) and gyroscope IMU sensor data. These omissions represent a limitation, as both green channel PPG and gyroscope IMU data could potentially improve motion artifact removal and physiological parameter estimation accuracy. Future iterations of the dataset should address these interconnected gaps to provide a more complete foundation for algorithm development.

\subsubsection{Limits of RingTool Benchmark}
\label{sec: discussion sub3.2}
Our current benchmark includes four common algorithm baselines, but does not encompass all their variants or a wider range of potential algorithms. Currently, our toolkit only supports the $\tau$-Ring dataset as the sole open-source ring-based physiological dataset, and we plan to expand compatibility with future open-source ring datasets as they become available. While we have demonstrated that incorporating accelerometer data improves heart rate estimation during motion scenarios in Table~\ref{tab: ablation_study}, our current implementation applies equal weights to multi-channel PPG and ACC inputs. In real-world applications, these weights likely need further optimization and dynamic adjustment based on different usage contexts and user characteristics. Due to computational resource constraints, we were unable to conduct systematic ablation studies on critical parameters such as time window lengths, filtering frequency bands, model hyperparameters, and optimal sensor fusion strategies. We believe more comprehensive algorithmic investigations could address these gaps, potentially yielding improved performance across different sensing scenarios and physiological parameters.

\subsubsection{Optimization and Deployment on Ring Devices}
\label{sec: discussion sub3.3}
Our benchmark establishes a foundation for ring-based physiological sensing but requires further optimization for practical implementation. We observed that different vital signs benefit from specific algorithmic approaches (e.g., ResNet for HR and peak detection for RR estimation as shown in Table~\ref{tab: all_performance}), suggesting that a multi-task learning framework could improve overall efficiency. Moreover, scene-adaptive algorithms with specialized processing paths for different activity states could bridge the performance gap between stationary and motion scenarios by better handling varying levels of motion artifacts. However, with current models ranging from 0.1M to 2M parameters, on-device implementation remains challenging. Future work should focus on model compression techniques such as pruning, knowledge distillation, and quantization to reduce computational requirements while maintaining accuracy. These optimizations would enable on-device physiological sensing on resource-constrained ring platforms with improved battery efficiency and privacy.

\section{Conclusion}
\label{sec:conclusion}

In this paper, we presented $\tau$-Ring, the first ring-based dataset designed for cardiovascular physiology sensing. $\tau$-Ring consists of infrared and red channel PPG signals with 3-axis ACC data collected from both reflective and transmissive optical paths. With 28.21 hours of raw data from 34 subjects, $\tau$-Ring encompasses seven activities with different motion artifacts and four ground truth labels. 
We evaluated three widely-used physics-based methods and four cutting-edge deep learning methods using our proposed RingTool toolkit. The results show our ring-based systems achieve better performance than commercial rings, with the best MAE of 5.18 BPM for HR estimation, 2.98 BPM for RR estimation, 3.22\% for SpO2 estimation, and 13.33/7.56 mmHg for SBP/DBP estimation. We further evaluated performance across different scenarios, demonstrating that our ring-based systems are reliable in stationary scenarios but face challenges in motion scenarios. Through ablation studies on the fusion of multi-channel input, we demonstrated that ACC fusion can potentially improve the performance of PPG-based systems.
We believe that $\tau$-Ring and RingTool will be valuable resources for the research community to advance the accuracy and reliability of cardiovascular physiology sensing.

%%
%% The acknowledgments section is defined using the "acks" environment
%% (and NOT an unnumbered section). This ensures the proper
%% identification of the section in the article metadata, and the
%% consistent spelling of the heading.
\begin{acks}

\end{acks}

%%
%% The next two lines define the bibliography style to be used, and
%% the bibliography file.
\bibliographystyle{ACM-Reference-Format}
\bibliography{refs}

%%
%% If your work has an appendix, this is the place to put it.

\end{document}